\documentclass[preprint2]{aastex}

\usepackage{graphicx}
\newcommand{\citegenitive}[1]{\citeauthor{#1}'s \citeyearpar{#1}}

\shorttitle{SFH of SMC field}
\shortauthors{Cignoni et al.}

\begin{document}


\title{Star formation history in two fields of the Small Magellanic
Cloud Bar\altaffilmark{1}}


\author{M. Cignoni\altaffilmark{2,3}, A. A. Cole\altaffilmark{4},
  M. Tosi\altaffilmark{3}, J. S. Gallagher\altaffilmark{6},
  E. Sabbi\altaffilmark{5}, J. Anderson\altaffilmark{5},
  E. K. Grebel\altaffilmark{7}, A. Nota\altaffilmark{5}}

\email{michele.cignoni@unibo.it}

    \altaffiltext{1}{Based on observations with the NASA/ESA Hubble
      Space Telescope, obtained at the Space Telescope Science
      Instutute, which is operated by AURA Inc., under NASAcontract
      NAS 5-26555. These observations are associated with program
      GO-10396.}  \altaffiltext{2}{Astronomy Dept. University of
      Bologna, Bologna, 40127, Italy)} \altaffiltext{3}{INAF Bologna
      Observatory, 40127, Italy}\altaffiltext{4}{School of Mathematics
      \& Physics, University of Tasmania, Private Bag 37, Hobart,
      Tasmania 7001, Australia } \altaffiltext{5}{STScI, Baltimore,
      MD, 21218, USA} \altaffiltext{6}{University of Wisconsin,
      Madison, WI, USA} \altaffiltext{7}{Astronomisches
      Rechen-Institut, Zentrum f\"ur Astronomie der Universit\"at
      Heidelberg, M\"onchhofstr.\ 12--14, 69120 Heidelberg, Germany }

\begin{abstract}

  The Bar is the most productive region of the Small Magellanic Cloud
  in terms of star formation but also the least studied one. In this
  paper we investigate the star formation history of two fields
  located in the SW and in the NE portion of the Bar using two
  independent and well tested procedures applied to the
  color-magnitude diagrams of their stellar populations resolved by
  means of deep HST photometry. We find that the Bar experienced a
  negligible star formation activity in the first few Gyr, followed by
  a dramatic enhancement from 6 to 4 Gyr ago and a nearly constant
  activity since then. The two examined fields differ both in the rate
  of star formation and in the ratio of recent over past activity, but
  share the very low level of initial activity and its sudden increase
  around 5 Gyr ago. The striking similarity between the timing of the
  enhancement and the timing of the major episode in the Large
  Magellanic Cloud is suggestive of a close encounter triggering star
  formation.

\end{abstract}





\section{INTRODUCTION}

 The Small Magellanic Cloud (SMC) is a fundamental laboratory to
  study the evolution of dwarf irregular galaxies (dIrr's). The SMC is
  the closest member of this class of systems, has a current
  metallicity (Z$\,\simeq 0.004$ as derived from HII regions and young
  stars) similar to that of the majority of dIrr's and a mass
  \citep[between 1 and $5 \times 10^9\, M_{\odot}$, e.g. ][and
  references therein]{kallivayalil06} at the upper limit of their
  range. Moreover the SMC is a member of a triple system, a
  circumstance that favors studying the modulation of the star
  formation activities driven by interactions.

  In order to derive the detailed, spatially-resolved star formation
  history (SFH) of the SMC we have started an international long-term
  project to study the evolution of the SMC in space and time
  \citep[see][]{tosi08}. Our strategy is to achieve high photometric
depth and spectroscopic resolving power over a large field of view by
combining datasets from the ground and space.  We are using the Hubble
Space Telescope (HST), the Very Large Telescope (VLT), and the VLT
Survey Telescope (VST) to observe a large sample of field stars and
clusters across the SMC. These data will allow us to constrain the
global SFH as well as the existence of chemical and age gradients.

For the cluster analysis, we have already presented deep photometry
with HST's Advanced Camera for Surveys (ACS) of seven intermediate-age
and old populous clusters (\citealt{glatt08a, glatt08b, glatt09,
  glatt11}). In combination with our VLT data we find a complex
age-metallicity relation for these clusters with a considerable spread
in metallicity at any given age (see e.g. \citealt{glatt08b}).

\begin{figure*}
\centering 
\includegraphics[width=9cm]{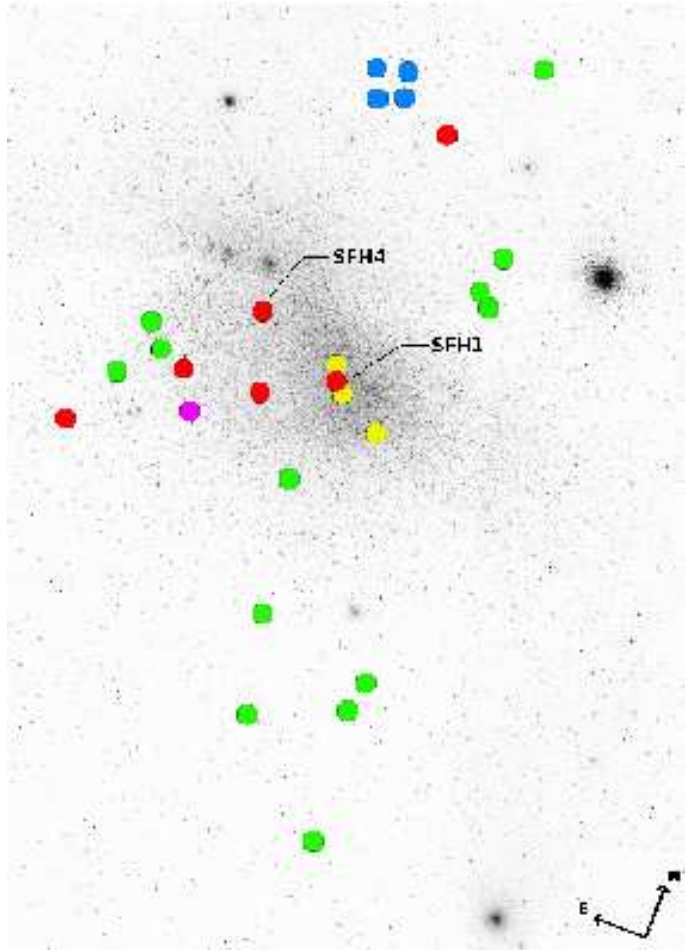}
\caption{Spatial distribution of the six observed fields (red symbols)
  together with the observations from \citet{dolphin01} (blue
  symbols), \citet{mccumber05} (magenta symbol), \citet{noel07} (green
  symbols), \cite{chiosi07} (yellow symbols), superimposed on the DSS
  image of the SMC.}
\label{fields} 
\end{figure*}

Concerning the SMC field analysis, our plan is to have Color-Magnitude
diagrams (CMDs) several magnitudes fainter than the oldest
main-sequence (MS) turn-off (TO) for the entire galaxy. To this
purpose we have observed six fields with the ACS \citep{sabbi09},
sampling regions characterized by different stellar and gas densities
in the SMC Bar, in the Wing in the direction of the LMC, and in the
outskirts (see Fig. \ref{fields}). A preliminary SFH analysis of such
fields has been presented in \citet{sabbi09}. We also have an ongoing
Guaranteed Time Observation program at the VST \citep{ripepi06}
designed to cover with deep photometry the whole SMC and the Bridge
connecting it to the LMC. These CMDs will allow us to infer for the
first time the SFH of the whole SMC over the entire Hubble time,
covering a much larger area with considerably better image quality
than \citegenitive{zaritsky2002} data. We will derive the SFHs from
the CMDs using the synthetic CMD technique \citep[see e.g.][and
references therein]{tolstoy09,cignoni10}.
SFHs of some SMC fields have been derived and presented by other
authors, based on ground-based observations or HST studies of small
individual regions (see Fig. \ref{fields}). \citet{harris04} derived
the SFH of the SMC over $4\degr \times 4.5\degr$ to a depth of $V\la
21$ using the Magellanic Cloud Photometric Survey (MCPS) UBVI catalog
\citet{zaritsky2002}. This is currently the most spatially extended
study of the galaxy, but does not reach the oldest MSTO. The most
comprehensive study of the old population of the SMC to date was
carried out by \cite{haschke12} using RR Lyrae stars from the Optical
Gravitational Lensing Experiment (OGLE-III; \citealt{udalski2008}).
They find a uniform metallicity distribution across the SMC with a
spread of more than 1 dex in [Fe/H]. \citet{dolphin01} analyzed the
stellar content of the SMC halo, in a region close to the globular
cluster NGC~121, using both HST Wide Field Planetary Camera 2 (WFPC2)
and ground based data.  Again with WFPC2 \citet{mccumber05} studied
the stellar content of a small portion of the SMC Wing.
\cite{chiosi07} derived the SFH in the vicinity of a few SMC
clusters. Finally, \citet{noel07} and \cite{noel09} presented a deep
ground-based study of 12 fields of the SMC, avoiding the densest
regions, because of their high crowding conditions.

In this paper we present the SFH of SFH1 and SFH4, the two most
central fields of the six SMC regions we observed with HST/ACS
\citep{sabbi09}.  The apparent distances from the SMC optical center
are about 24$^{\prime}$ and $1^{\circ}\,52^{\prime}$
respectively. SFH1 is located in the SW portion of the SMC Bar, where
the stellar density, gas and dust contents are highest, while SFH4 is
located to the NE of the SMC center, at 24$^{\prime}$ south of
NGC~346, the most active star-forming region in the SMC.

For a better assessment of the intrinsic theoretical uncertainties,
the SFH is derived using two completely independent procedures for the
application of the synthetic CMD method. We compare the two methods
here and discuss the corresponding results. The other ACS fields
observed by us and described by \cite{sabbi09} are being treated in
the same way, and their SFH will be presented in a forthcoming paper.

We briefly describe our data in Section 2. The two procedures for the
SFH derivation are summarized in Section 3, together with the results
of their application to SFH1 and SFH4. Similarities and differences
between the resulting SFHs are discussed in Section 4, while in
Section 5 we compare our findings with previous literature. Concluding
remarks follow in Section 6.

\section{Photometry and CMDs of SFH1 and SFH4}

The data for our six SMC fields were acquired with the ACS Wide Field
Channel between November 2005 and January 2006 (GO-10396;
P.I. Gallagher) with the F555W and F814W filters. The data reduction
was performed with the program img2xym\_WFC.09x10 \citep{anderson06},
and the resulting magnitudes were calibrated in the Vegamag
photometric system using \citet{sirianni05} recipes. For sake of
simplicity, from now on we will refer to the $m_{F555W}$ and
$m_{F814W}$ magnitudes calibrated in the Vegamag system as V and I,
respectively.

Extensive artificial star experiments were performed to test the level
of completeness and the photometric errors of the data. They followed
the approach described in \citet{anderson08}, and the artificial stars
were searched for with exactly the same procedure adopted for the real
stars. We considered an artificial star as recovered if its input and
output positions agree to within 0.5 pixels and the fluxes agree to
within 0.75 mag. As done in the photometric analysis we also required
that each star was found in at least three exposures with a positional
error $<0.1$ pixels per filter.

Details on both the data reduction and the artificial star tests can
be found in \cite{sabbi09}.
\begin{figure*}
\centering
\includegraphics[width=15cm]{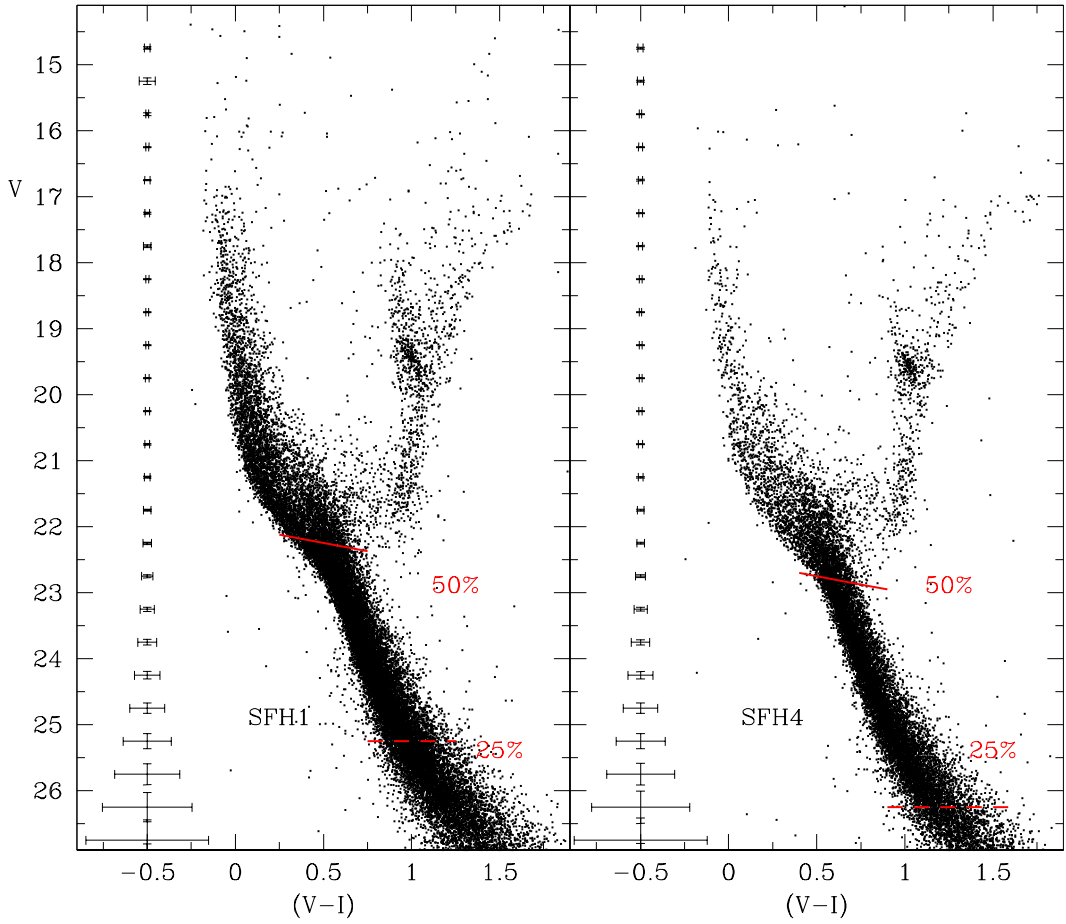}
\caption{CMDs of the SFH1 (left panel) and SFH4 (right panel) field
  observed with the ACS/WFC. The solid and the dashed red lines
  indicate the 50\% and 25\% levels of completeness, respectively.
  Formal errors on the estimated photometry are shown on the left side
  of each CMD (see text for details).}
\label{cmdobs} 
\end{figure*}

The final SFH1 and SFH4 catalogs contain about 29200 and 17300 stars
respectively, and the corresponding CMDs are shown in
Fig.\ref{cmdobs}, where the photometric errors are also plotted
\footnote{To be conservative, the plotted error bars correspond at
  each magnitude level to the larger value of the error resulting from
  the photometric package and from the artificial star tests. }. These
CMDs reach almost four magnitudes fainter than the oldest MSTO, thus
allowing to study the evolution of the regions over the whole Hubble
time. The sub-giant branch (SGB) is well populated and its brightness
extension, much larger than the photometric error at that magnitude
level, suggests a prolonged star formation between 3 and 12 Gyr
  ago. The red giant branch (RGB) and the red clump (RC) also are well
  populated. In addition to these intermediate-age and old components,
  both CMDs show a MS blue plume and a blue loop (BL) sequence typical
  of late-type dwarf galaxies, corresponding to young high- and
  intermediate-mass stars.

Finally, the CMDs do not show any significant population of stars on
the right of the lower MS, the CMD locus occupied by pre-main sequence
(PMS) stars. This, in turn, suggests a negligible activity in the last
50 Myr (the average time spent by a solar-like star in PMS; see, e.g.,
\citealt{cignoni10b}) compared to more active regions of the SMC like
NGC~346 or NGC~602 (see e.g. \citealt{nota2006, cignoni09, carlson11,
  cignoni11}).

\section{SFH derivation: the two synthetic CMD approaches}

The SFH of SFH1 and SFH4 has been derived with the synthetic CMD
method. As recently reviewed by \cite{tolstoy09} and \cite{cignoni10},
this approach was first applied to nearby resolved galaxies twenty
years ago \citep{tosi91} and is now recognized as a powerful tool to
disentangle the complex history of resolved galaxies. To both fields
we have applied it following two independent procedures: Cole's
\citep[e.g.][]{cole07} and a combination of Cignoni's
\citep[e.g.][]{cignoni06} and the Bologna
\citep[e.g.][]{tosi91,greggio98,angeretti05} codes. These tecniques
supersede the classical isochrone fitting approach, allowing us to
explore a much wider parameter space and to incorporate statistical
and observational uncertainties. Nonetheless their use is limited by
the reliability of the adopted stellar tracks and color
transformations, as well as the nature of the IMF (e.g., $\S$4.1) and
quality of the data. The recovered history is the best of all
possibilities, but this does not necessarily imply that it is the
actual solution nor that it is unique.

The procedures used here have already been tested in comparisons with
each other and similar methods.  Both groups participated in the
Coimbra experiment in 2001 \citep[see][and references
therein]{skillman02}, Cignoni's method has been applied to the
derivation of the Solar Neighborhood SFH \citep{cignoni06,cignoni06b}
and in combination with the Bologna code to the analysis of the SMC
young clusters NGC~346 \citep{cignoni10b,cignoni11} and NGC~602
\citep{cignoni09}, Cole's method has been compared with two different
codes in the analysis of the dwarf irregular IC~1613
\citep{skillman03} and the Cetus dwarf spheroidal \citep{monelli10}.
Since the procedures are in continuous evolution, to improve both the
reliability of the results and the efficiency of the computations, it
is useful to perform further comparisons on deep and tight CMDs such
as those of SFH1 and SFH4.

In the following we briefly summarize commonalities and differences of
the two methods.  In both cases the synthetic CMDs have been built
using the results of the artificial star tests mentioned above to
assign photometric errors and incompleteness to the synthetic
stars. 


\subsection{Bologna procedure}

The Bologna approach involves the comparison of the observed CMD with
a library of synthetic CMDs computed with different values of
metallicity, initial mass function (IMF), binary fraction, distance
modulus and reddening. Models have been calculated with the latest
Padova stellar models (\citealt{bertelli08}, \citealt{bertelli09}) for
masses between the hydrogen burning limit and $20\,
M_{\odot}$. Theoretical temperature and luminosity are transformed to
the observational plane by means of the relations obtained by
\cite{origlia00} for the HST Vegamag photometric system.  The
synthetic CMD is created following a classical Monte Carlo procedure:
1) stellar masses and ages are randomly extracted from a time
independent IMF and a star formation law; 2) the stellar tracks are
interpolated deriving the absolute photometry for the synthetic
population; 3) corrections for the distance modulus and the foreground
reddening are applied. Then the synthetic CMD is degraded to match
both the completeness profile and the photometric error properties as
derived from extensive artificial star tests (\citealt{sabbi09}). To
be conservative, we have limited our fitting procedure to stars
brighter than $V=23$, whose completeness is better than 40\% and
photometric errors smaller than 0.05 mag.

The full SFH of a galaxy can be a complex function of time. To make
the problem manageable, we were forced to limit the range of parameter
space that our models cover on the basis of previous results and
indications from our data. In order to reduce the computational time,
the star formation rate is parametrized as a set of constant values
over adjacent temporal steps: a generic CMD is a linear combination of
chosen basis CMDs, where each basis is a Monte Carlo extraction from a
step star formation episode. The duration of each step is chosen in
relation to the evolutionary timescale of the average stellar mass of
the step, ranging from 50 Myr (approximately the MS timescale for an
$8\,M_{\odot}$ star) to 1 Gyr (as representative of the theoretical
precision at 10 Gyr, i.e. 10\%) for ages above 3 Gyr. The star
formation rate is constant within each step. In this work we deal with
90 basis CMDs, each representing the synthetic photometry of one of 18
age bins and one of five metallicities ($Z=0.008$, $Z=0.004$,
$Z=0.002$, $Z=0.001$, $Z=0.0004$). In order to reduce the Poisson
fluctuations, all partial CMDs are generated with more than $10^6$
solar masses.

Within the framework of the adopted stellar tracks and atmosphere
models, the most likely solution to the underlying SFH is the one
which minimizes the differences between data and synthetic star counts
over strategic regions of the CMD. The degree of likelihood is
assessed through a $\chi^2$ function of the residuals. Our experience
suggests that the performance of such minimization is very sensitive
to the adopted CMD binning scheme. Both fine and coarse grids offer
advantages, as well as disadvantages (see
e.g. \citealt{cignoni06}). Along the MS, a fine grid is mandatory to
study the old stellar generations, since they are tightly packed
together in the CMD. However, such a solution would pay the penalty of
under-weighting the star counts along the upper MS which is Poisson
dominated. Vice versa, a coarse grid would be more adequate to map the
recent activity, but it would allow a worse resolution at early
times. The situation is even worse with the evolved stars, where the
theoretical uncertainties are typically larger than for MS stars: the
CMD position of horizontal branch (HB) and RC stars is affected by a
complex interplay of age, metallicity, mass loss and helium content
(see e.g. \citealt{castellani00}); not only the color but also the
shape of the RGB are strongly affected by the color transformations
from the theoretical to the observational plane and by the mixing
length parameter (in a way which depends on the used wavelength); the
BL morphology is affected by the He burning cross sections (especially
the $^{12}$C($\alpha$,$\gamma$)$^{16}$O) as well as by the
efficiencies of external convection, overshooting and mass
loss. Nevertheless, the study of evolved stars can prevent the
ambiguities in models restricted to MS star counts, for example, when
photometric errors and incompleteness hinder the possibility to trace
the old activity by means of low mass MS stars.  Given these issues,
the problem of the grid choice can be solved only by using a variable
grid spacing. Several tests were conducted for finding the optimal
configuration. 
\begin{figure*}[t]
\centering
\includegraphics[width=8cm]{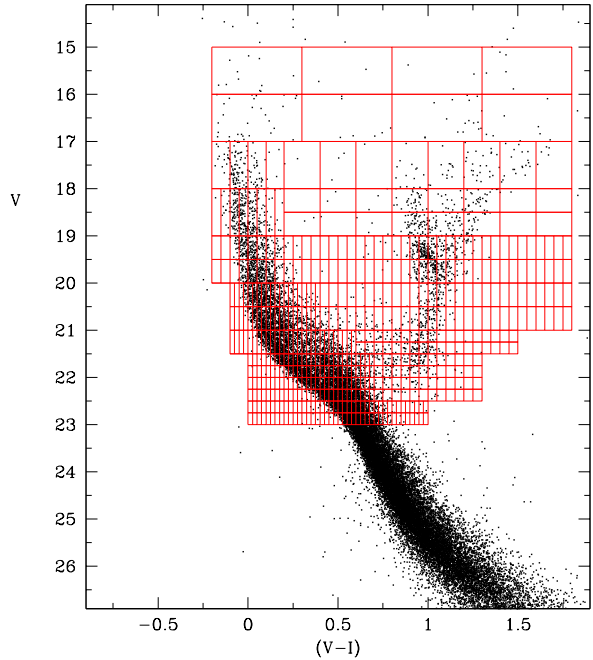}
\caption{CMD grid used to derive the SFH via the Bologna procedure.}
\label{grid} 
\end{figure*}
 Figure \ref{grid} shows our best scheme. To capture all
the information contained along the MS, the grid spacing shrinks with
luminosity, balancing the longer evolutionary times of lower mass
stars and, at the same time, providing sufficiently good statistics in
the upper MS. Towards the red part of the CMD the average grid spacing
is coarser: this helps to take the uncertainties into account, while
preserving the number of stars in specific phases.

Finally, the $\chi^2$ is minimized by means of a downhill simplex
algorithm. In order to avoid local minima, the simplex is re-started
from thousands of initial random positions. A bootstrap method is used
to assess the effect of random errors. The search of the best SFH is
repeated for each bootstrapped data set, producing a distribution of
best solutions. The errorbars on the final SFH represent one standard
deviation using 100 bootstraps.

With this procedure we have analyzed the CMD of SFH1 \& SFH4. As a
first step we have derived the SFH assuming a two-exponent power law
IMF (see \citealt{kroupa2001}) with exponent s$=$2.3, close to
Salpeter's 1955\footnote{Salpeter's IMF has exponent s$=$2.35 and the
  form $\int^{+\infty}_{0}m\,\phi(m)\,dm=1$ and $\phi(m)=m^{-s}$.} for
masses above $0.6\,M_{\odot}$ and $s=1.3$ below, and a binary fraction
(with primary and secondary mass extracted from the same IMF) of 30\%
(this parameter has no substantial impact on the fit quality). To
reduce the parameter space, only three metallicities, $Z=0.004$,
$Z=0.002$ and $Z=0.001$, have been adopted. The distance modulus and
the foreground reddening were allowed to vary freely.  The solutions
with the lowest reduced $\chi^2$ are listed as SFH1-A and SFH4-A for
the respective fields.

In order to test how much the details of the solution depend on the
allowed metallicity range, we calculated a second solution where the
Padova tracks with Z = 0.0004 and Z = 0.008 were also included.  These
solutions are presented as SFH1-B and SFH4-B. These are generally
consistent with the solutions obtained with the restricted metallicity
range but provide a somewhat better match to the observed CMD as
described in \S\ref{sec-sfh1} below.

\subsection{Simulated annealing procedure (Cole's method)}

The code developed by Cole has many features in common with the
Bologna code, but also incorporates some differences in the treatment
of the Hess diagram data and in the calculation of the merit function
that measures the relative likelihood of various solutions.  Because
the errors in extracting detailed information from a CMD are dominated
by systematic effects, the application of multiple fit procedures to
the same data can provide important insight into which features of the
SFH are robust and which may be artifacts.

Cole's SFH-fitting code begins with a set of theoretical isochrones
interpolated to a fine grid of age and metallicity in order to create
a synthetic CMD with no gaps.  The most recent isochrones from the
Padova models are used; all calculations are made in the HST Vegamag
system to minimize the possibility of introducing errors in
transformation equations.  The synthetic CMDs are binned in age and
metallicity (Z) to increase the speed of computing and to avoid
``overfitting'' noise in the CMDs.  We begin with bins that are evenly
spaced by $\approx$0.10 in $\Delta$log(age) and 0.20 in
$\Delta$log(Z), and merge adjacent bins when the noise level of the
solutions indicates there is insufficient information content in the
CMD to resolve the fine age bins.

The CMD is divided into a regular grid of color-magnitude cells, and
the expectation value of the number of stars in each cell for a SFR of
1 M$_{\odot}$ yr$^{-1}$ is calculated from the isochrones.  There are
several parameters that are taken as fixed constants during the
solution.  These include the IMF, fraction of binaries and mass ratio
distribution function, and the distance modulus and reddening.  The
adopted IMF is from \citet{chabrier03}, and the binary fraction and
mass ratios are parametrized based on \citet{duquennoy91} and
\citet{mazeh92}.  In this prescription, 35\% of stars are single and
the rest are binary.  The binaries are divided into ``wide'' and
``close'' binaries in a 3:1 ratio; the secondaries in the wide systems
are drawn from the same IMF as the primaries, but in the close systems
the secondary masses are drawn from a flat IMF.  The distances and
reddenings are initially constrained to the values given in
\citet{sabbi09}, but are varied if the resulting synthetic CMDs are
mismatched to the data.

No age-metallicity relation is explicitly assumed, but a range of
metallicities at each age is allowed, constrained by the color range
of the data.  In some cases, there is little leverage in the CMD to
constrain the metallicity, so outside information is used to choose
the metallicity.  For example, the V$-$I color of the upper MS is not
strongly metallicity-dependent, so we adopt a metallicity of Z = 0.004
based on HII region and Cepheid metallicities for the youngest stellar
populations in the SMC Bar.  For stars with ages on the order of
$\approx$1--7$\times$10$^8$ yr, this metallicity also gives a good
match between the colors of the blue supergiant stars in SFH1 and the
Padova models.  For older ages, ranges of metallicity at each age are
allowed.

A synthetic CMD is constructed by convolving the weighted,
color-magnitude binned isochrones with the color and magnitude errors
and incompleteness functions determined by artificial star tests.  In
the fit process, linear combinations of the individual synthetic CMDs
are added to find the composite CMD that best matches the data.
Because many cells in the Hess diagram are empty or contain few stars,
a maximum likelihood test based on the Poisson distribution must be
used instead of a $\chi^2$ statistic \citep{cash79}.  A direct search
of parameter space is infeasible because there may be dozens of age
and metallicity bins in the solution; additionally, effects such as
age-metallicity degeneracy can easily produce a large number of false,
local maxima in the likelihood space.  Because of this we use a
simulated annealing approach, in which a simplex of initial guesses at
the SFH is randomly perturbed and the changes are rejected with some
probability if they worsen the fit.  The star formation rates are
transformed according to an arcsin function prior to fitting in order
to prevent negative star formation rates from being considered.

Errorbars on the SFH are calculated by testing each age-metallicity
bin of the best-fit solution in turn.  The SFR in the bin under
consideration is forced away from the optimal solution and the fitting
procedure is redone, with the tested bin held fixed.  The 1$\sigma$
error on the SFR of the bin is taken to be the value beyond which no
fit can be found that is within 1$\sigma$ of the global best fit.
Because the total number of stars is fixed, a deficiency of stars in
one age bin can often be partially compensated by increasing the SFR
in nearby bins; this means that errorbars are fairly large and the SFR
in adjacent bins can be strongly anticorrelated.  Because there are
always unmodelled populations (including Galactic foregrounds,
background galaxies, and simple bad data) and there may be poorly-fit
stellar sequences (e.g., the colors of RGB stars), the overall fit
quality as measured by the equivalent of a $\chi^2$ statistic is
usually found to be quite poor; only the relative likelihoods are of
any meaning.  The most likely SFH to match the observed CMD is driven
by the most populous cells in the CMD; because these are frequently
near the faint end of the data where incompleteness is high, it is
absolutely essential to have a very accurate model of the
incompleteness to obtain robust results.

The ACS data are of high quality, but incompleteness sets in at a
relatively bright level.  The large number of stars and low average
error leads us to use magnitude bins 0.08 mag high and color bins 0.04
mag wide.  In order to avoid systematics due to the incompleteness, we
restricted the comparison to magnitudes 15 $\leq$ I $\leq$ 22.
Because there are significant uncertainties in modeling the colors of
RGB and RC stars, we restricted our fits to the MS and SGB.  We
considered a set of isochrones binned by 0.10 in log(age) from 4~Myr
$\leq$ age $\leq$ 13.5~Gyr (6.60 $\leq$ log(age) $\leq$ 10.13); below
log(age) = 8.60 the decreasing number of bright tracer stars led us to
double the size of the age bins.  We considered metallicities of Z =
0.00015, 0.0004, 0.0006, 0.001, 0.0015, 0.0024, and 0.004.

We begin with the distance moduli and reddenings from \citet{sabbi09},
but found that these needed to be modified in order to obtain the best
match to the data with the Padova isochrones.  For SFH1 we found the
best distance modulus (m$-$M)$_0$ = 18.83 and reddening E(B$-$V) =
0.08, while in SFH4 we used 18.85 and 0.12.  In both fields, the
values are within 1$\sigma$ of the initial guesses.  In SFH1 we found
that the model upper MS was significantly narrower and bluer than the
data with the adopted reddening value, so we were forced to introduce
an extra reddening component.  We adopted the simple expedient that
all populations younger than log(age) = 7.60 were subject to double
the mean reddening of the field, while stars with 7.60 $\leq$ log(age)
$<$ 8.00 were assigned a reddening halfway between the younger and
older stars. Differential reddening in the central SMC has previously
been observed by \citet{zaritsky2002}, who used multicolor photometry
to derive line of sight extinction corrections to individual bright
SMC stars.  They found a mean A$_V$ = 0.18 mag for stars with 5500 K
$\leq$ T$_{eff}$ $\leq$ 6500 K, and A$_V$ = 0.46 mag for stars with
T$_{eff}$ $\geq$ 12000 K; these numbers are in reasonable agreement
with the values adopted here for SFH1.  \citet{zaritsky2002} discuss
the physical reality of their derived reddening distributions and
conclude that small scale reddening variations may reasonably be
attributed to residual gas and dust in star-forming regions.  The vast
majority of old, cool stars will not be seen through star-forming
regions, and so the small highly-reddened tail of the old population
will not strongly influence the mean reddening.  Because the amount of
differential reddening has high spatial frequency, the numerical
agreement (as in SFH1) or lack thereof (as in SFH4) with the
conclusions of \citet{zaritsky2002} must be considered fortuitous.

Cole's best-fit solutions are presented here as SFH1-C and SFH4-C
respectively for the two fields.

\section{Results: SFH of two fields in the SMC Bar}
\begin{table*}[t]
\scriptsize
\begin{center}
\caption{Summary of SFH Solution Parameters$^a$}\label{tab-models}
\begin{tabular}{lccccccc}
\hline
Solution & Method & (m$-$M)$_0$ & E(B$-$V) & IMF & CMD binning & Metallicities \\
         &        &    (mag)      &  (mag)   &     & (color$\times$mag) & (Z$\times$10$^3$) \\
\hline
Field SFH1: &      &       &      &      &           &         \\
SFH1-A   & Bologna & 18.77 & 0.11 & Kroupa (2001)   &  variable$^b$ & 1, 2, 4 \\
SFH1-B   & Bologna &       &      &                 &   & 0.4, 1, 2, 4, 8 \\
SFH1-C   & Cole    & 18.83 & 0.08$^c$ & Chabrier (2003) &  0.04$\times$0.08 & 0.15, 0.4, 0.6, 1.0, 1.5, 2.4, 4.0\\
\hline
Field SFH4: &      &       &      &      &          &         \\
SFH4-A   & Bologna & 18.80 & 0.11 & Kroupa (2001)   &  variable$^b$ & 1, 2, 4\\
SFH4-B   & Bologna &       &      &      &          & 0.4, 1, 2, 4, 8\\
SFH4-C   & Cole    & 18.85 & 0.12 & Chabrier (2003) &  0.04$\times$0.08 & 0.15, 0.4, 0.6, 1.0, 1.5, 2.4, 4.0\\
\hline
\end{tabular}
\medskip\\
\end{center}
$^a$All models based on the Padova isochrone set; see text for details.\\
$^b$See Figure~\ref{grid}.\\
$^c$Differential reddening assumed for stars younger than 100~Myr.
\end{table*}
The properties of the three SFH solutions for each field are summarized
in Table~\ref{tab-models}.  While the two procedures use the same photometry
and isochrones, they differ in a number of important respects.  The application
of two completely independent codes to derive the SFH allows us to identify
the most model-independent features of the SFH, gives insight into the areas
of the CMD that are driving the fitting procedures to the results they give,
and gives an estimate of the systematic uncertainties in the fits.

\subsection{SFH1 field \label{sec-sfh1}}
\begin{figure*}[t]
\centering
\includegraphics[width=8cm]{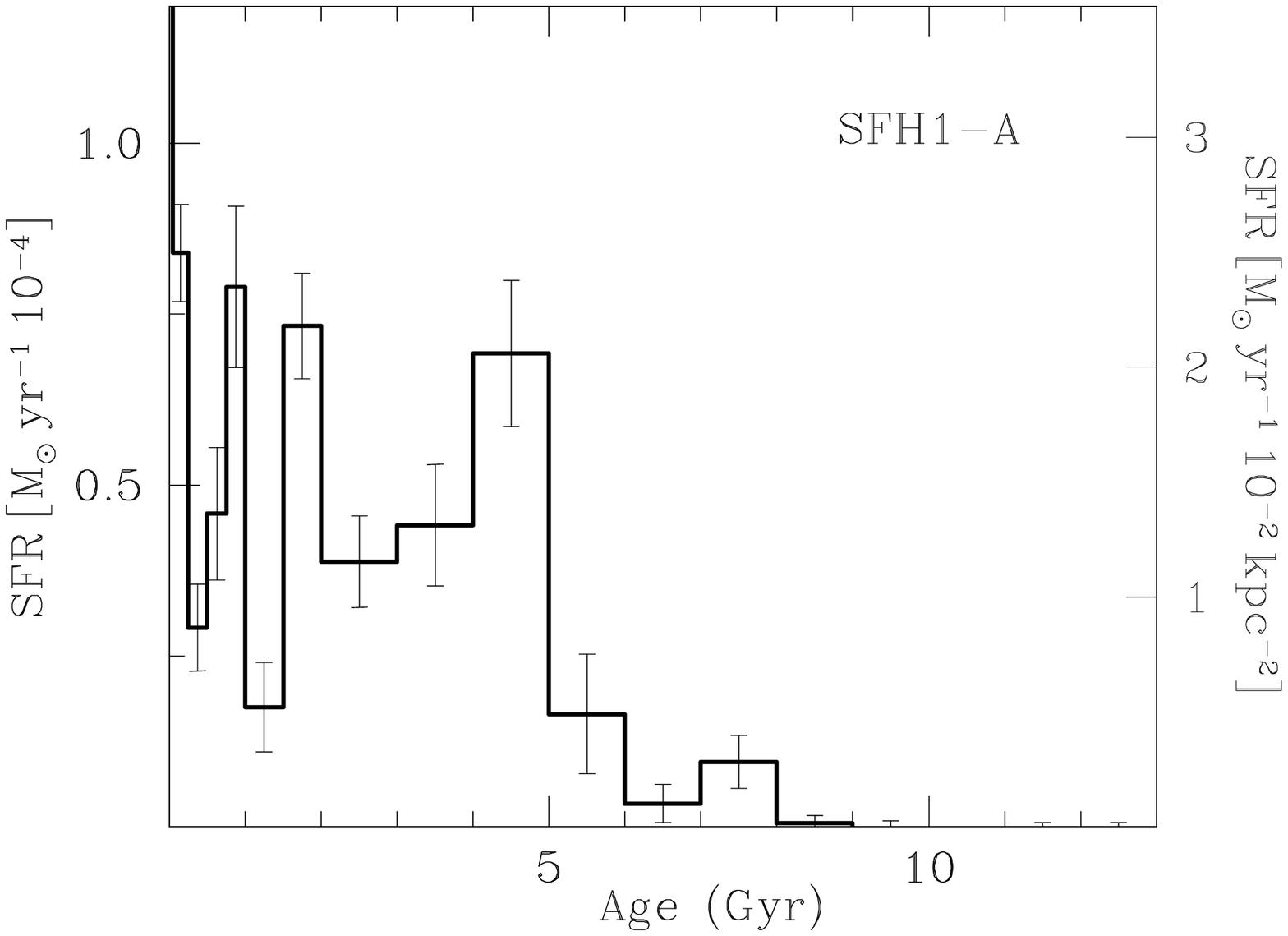}
\caption{SFH for the SFH1 field, obtained with a restricted metallicity
  range (solution SFH1-A). The star-formation 
  rate is given both in $M_{\odot}\,yr^{-1}$ (left ordinate) and per
  unit area ($M_{\odot}\,yr^{-1}\,kpc^{-2}$, in the right ordinate). }
\label{sfh} 
\end{figure*}
\begin{figure*}
\centering
\includegraphics[width=14cm]{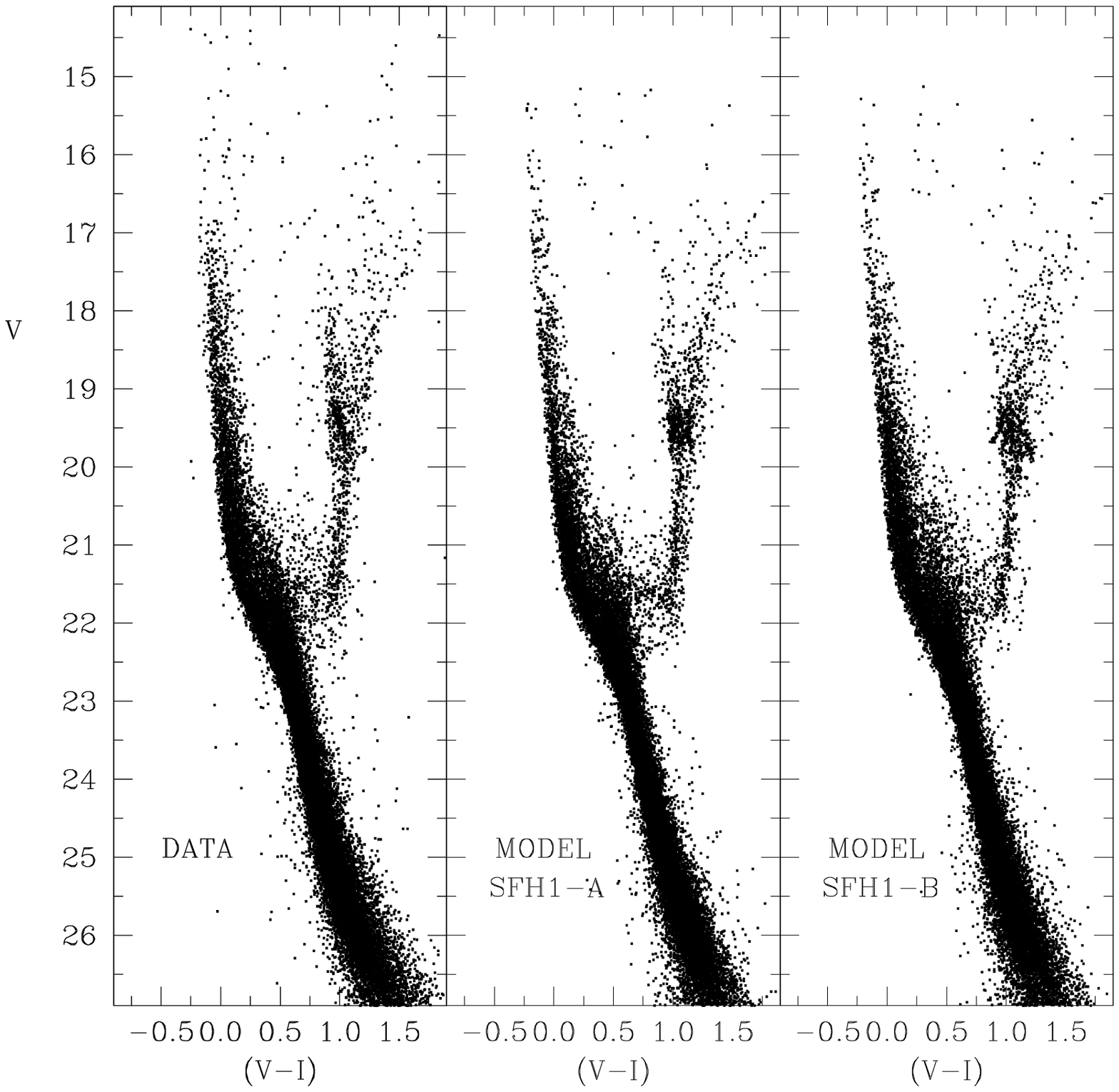}
\caption{Synthetic CMD for the solution SFH1-A (middle panel) and
  SFH1-B (right panel) compared with the observational CMD (left
  panel).}
\label{cmd_best} 
\end{figure*}

The SFH1-A (restricted metallicity) solution which gives the lowest
($2.6$) reduced $\chi^2$ is shown in Fig. \ref{sfh} (error bars
represent 1-$\sigma$ uncertainty), while the corresponding synthetic
CMD is shown in the middle panel of Fig. \ref{cmd_best}. The left
panel of the same figure shows the observational CMD for a direct
comparison. Our best distance modulus and reddening are
$(m-M)_0=18.77$ and $E(B-V)=0.11$, respectively. According to this
solution, the SFR has been extremely low during the earliest 6-7 Gyr,
with a strong and rapid increase around 5 Gyr ago. From then on, the
star formation activity has been typically gasping, with ups and downs
with respect to an average almost constant rate.
\begin{figure*}[t]
\centering
 \includegraphics[width=12cm]{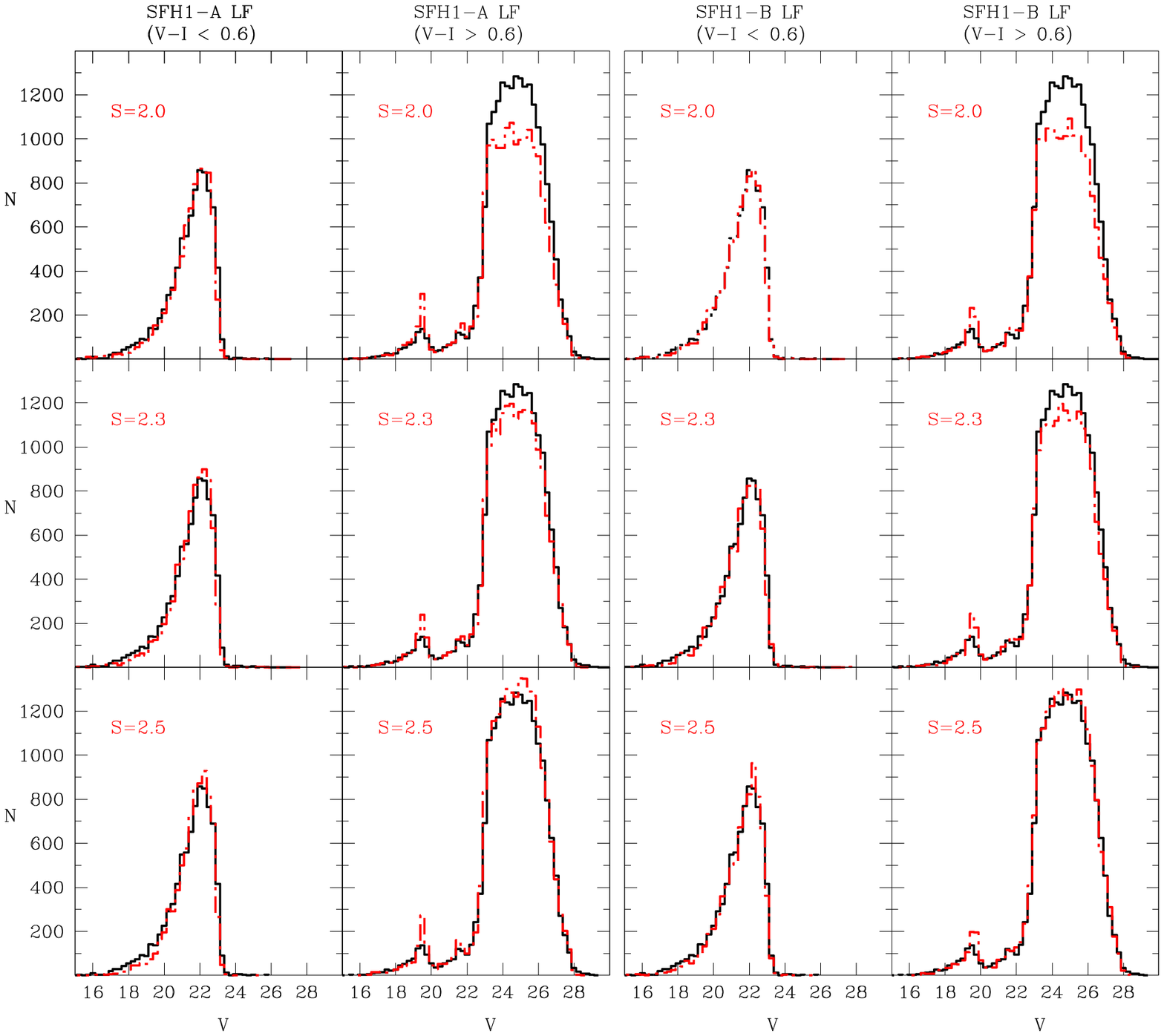}
 \caption{ SFH1: Predicted (red dot-dashed histograms) and
   observational (black solid histograms) LFs for stars bluer (first
   and third columns) and redder (second and fourth columns) than
   $V-I=0.6$ for the model solutions SFH1-A (first two columns) and
   SFH1-B (third and fourth colums). Models in the top, middle and
   bottom panels are computed with IMF exponents above
   $0.6\,M_{\odot}$ $s=$ 2.0, 2.3, 2.5, respectively. }
\label{LFs} 
\end{figure*}

In the middle row of Figure \ref{LFs} we plot the luminosity function
(LF) of blue and red stars (first and second column from the left,
respectively) in the observed CMD and in the one resulting from
SFH1-A. The top and bottom panels show for comparison the result when
the SFH is searched using IMF exponents $s=$ 2.0 and 2.5,
respectively, for stars above $0.6\,M_{\odot}$.

Despite a somewhat high reduced $\chi^2$ 2.6, this model is successful
in describing the main features of the data.

However when the results are examined in detail a few noticeable
issues can be identified. First of all the synthetic upper MS is
sharper than the observational one. As already noted by
\cite{sabbi09}, such a broadening may indicate an additional
absorption. One explanation could be that the SFH1 field is in the
main body of the SMC and the reddening material may be patchy and
cover not uniformly all lines of sight. Indeed, this is what is
observed in the extensive analysis of \cite{haschke11}, who suggests
for SFH1 a broader reddening distribution\footnote{Reddening values
  from the red clump method are available through the German
  Astrophysical Virtual Observatory interface at
  http://dc.zah.uni-heidelberg.de/mcx.}($\sigma_{E(V-I)}\approx 0.12$
mag) with respect to SFH4 ($\sigma_{E(V-I)}\approx 0.09$ mag). Yet
other possibilities are that a fraction of massive MS stars is
affected by rotation, which can lead to widened upper MSs in young
clusters (see e.g. \citealt{grebel96}), or a different SMC depth along
SFH1 and SFH4's line of sights.

Our model slightly but systematically underestimates the number of
blue massive stars brighter than $V\approx\,19$ by 20$-$30 percent
(see the middle panel of the first column in Fig. \ref{LFs}). A
flatter IMF could mitigate such discrepancy (see the top panel of the
first column in Figure \ref{LFs}), but the corresponding model would
underestimate the number of low mass stars (see the discrepancy
between $V=24$ and $V=26$ in the top panel of the second column in
Fig. \ref{LFs}).

As far as the evolved stars are concerned, the synthetic CMD matches
well both the SGB magnitude spread (that is a signature of the age
spread) and the RGB color dispersion below the RC luminosity. However,
there are several differences as well: 1) the predicted RC morphology
is irregularly shaped, while the observational RC is smooth and rather
elliptical. This difference may be partially explained by the coarse
metallicity resolution of our model, but also by a small amount of
differential reddening; 2) the synthetic CMD overestimates the number
of RC stars by about a factor of two; a steeper IMF could slightly
mitigate this mismatch (see the bottom panel of the second column in
Fig. \ref{LFs}) but at the expenses of the upper MS fit; 3) the RGB
stars brighter than the RC are too blue (even if the predicted counts
match exactly the observed ones). This suggests that our models are
too metal poor or, more likely, that the adopted color transformations
systematically fail near the RGB tip.

\begin{figure*}[t!]
\centering
 \includegraphics[width=16cm]{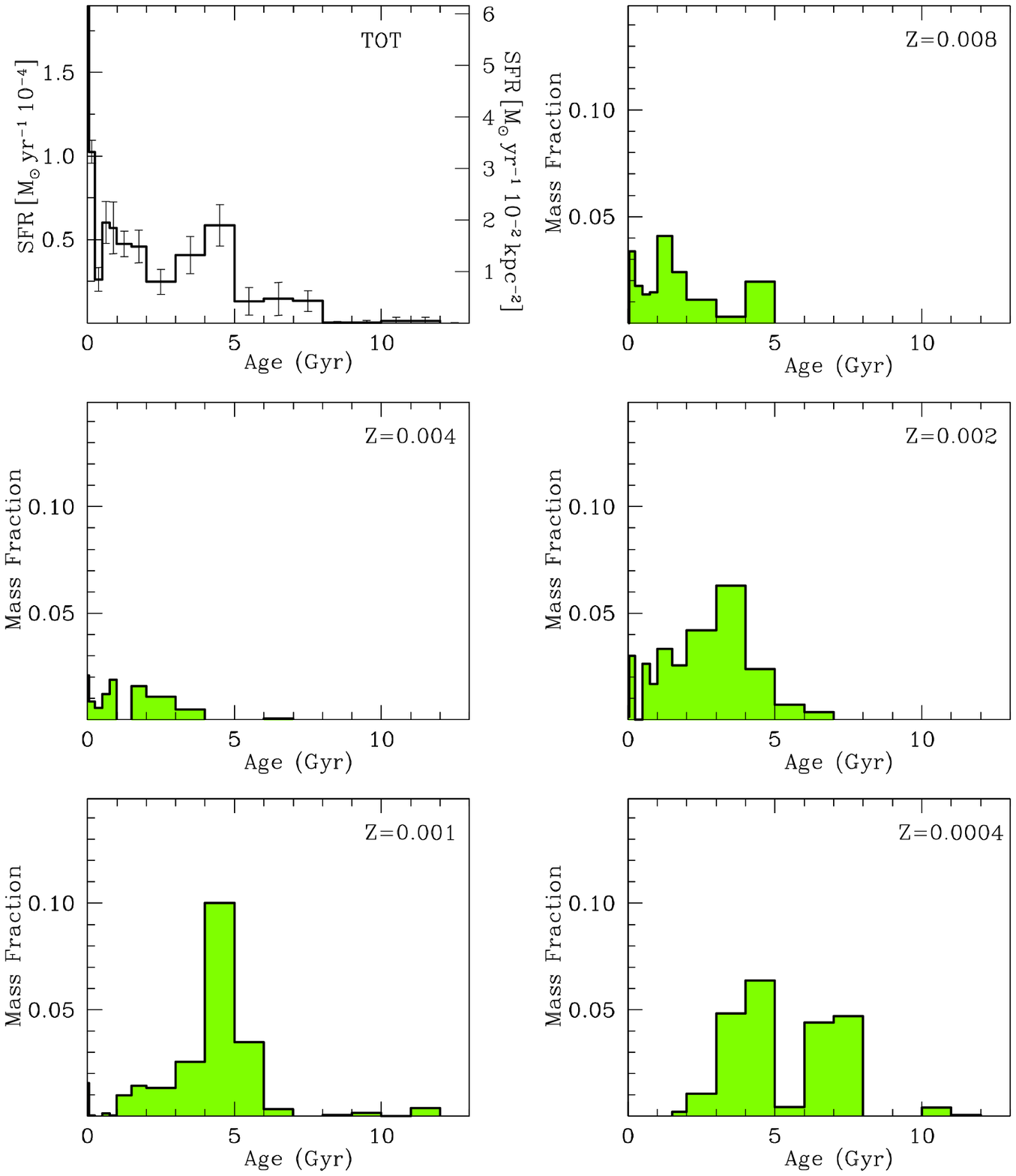}
\caption{SFH for the SFH1 field resulting from models which include
  also the metallicities $Z=0.008$ and $Z=0.0004$ (solution
  SFH1-B). The top-left panel shows the total SFH while the others
  display the contribution from each of the labeled metallicities.}
\label{SFH_008_0004} 
\end{figure*}

We investigate the effect of metallicity on the SFH found via the
Bologna procedure by adding the Padova tracks with $Z=0.008$ and
$Z=0.0004$ (solution SFH1-B). The top-left panel of
Fig. \ref{SFH_008_0004} shows the resulting SFH, while the other
panels of the same figure present the mass fraction contributed by
each metallicity.

\begin{figure*}[t!]
\centering
 \includegraphics[width=9cm]{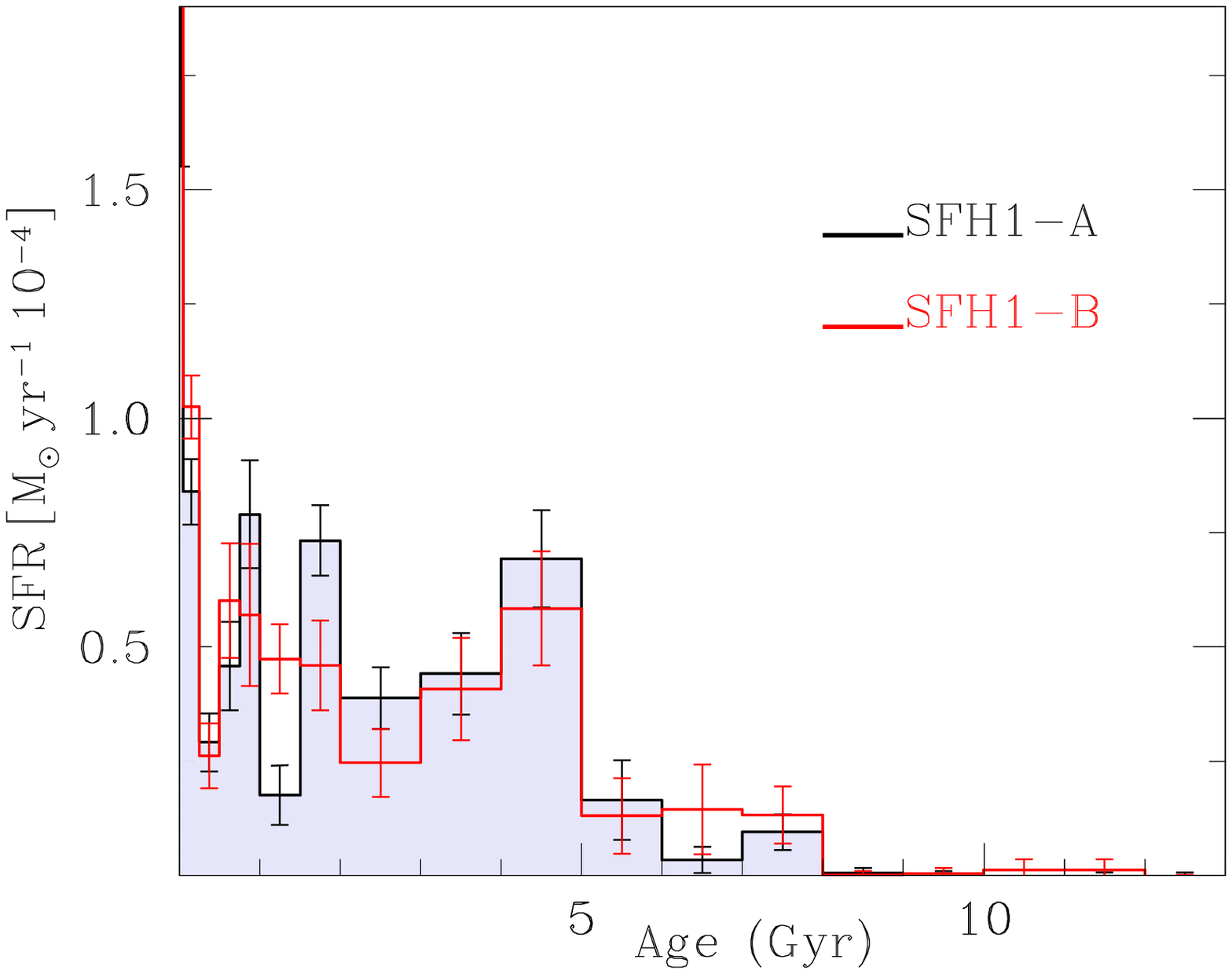}
\caption{Recovered SFHs for SFH1 using the Bologna procedure:
  SFH1-A vs.\ SFH1-B.  The latter is a better match to the data.}
\label{1_a_b} 
\end{figure*}

Overall, the solutions SFH1-A and SFH1-B are rather similar (see
Fig. \ref{1_a_b}), both being characterized by a long quiet period at
the oldest epochs. However, solution SFH1-B shows some different
features. First, the early activity is slightly higher, a direct
consequence of the lower initial metallicity; then, the activity
between 0.5 and 3 Gyr ago is smoother than in SFH1-A.

We find with additional test models that the epoch of the first peak
of star formation activity is progressively earlier for decreasing
values of the adopted lowest metallicity. We are thus confident that
our conclusion of a long almost quiescent period earlier than 5 Gyr
ago is robust, although we do see and predict some stars as old as
10-12 Gyr.

Concerning the recovered chemical history, we find that the largest
stellar mass fraction is produced at $Z=0.001$, while the contribution
from $Z=0.004$ is only a few percent. Generally, the metallicity
increases with time.
 

From the point of view of the fit quality, the new solution improves
the CMD match, yielding a reduced $\chi^2$ of about 2.3.  The
differences are visible both in the synthetic CMD of Fig.
\ref{cmd_best} (right panel) and in the LF (third and fourth columns)
of Fig.  \ref{LFs}.  Model SFH1-B reproduces the upper MS star counts
and morphology better than model SFH1-A (compare the middle panel
  of the third column with the middle panel of the first column in
  Fig.\ref{LFs}), but some problems still affect the simulation: 1)
the synthetic CMD shows a hint of a red HB, which is not observed in
the data; 2) the number of BL stars is still under-predicted; 3) the
RGB above the RC is too blue; 4) the predicted RC is still
overpopulated and irregularly shaped.

Given these results, especially the improvement along the upper MS,
model SFH1-B is globally better than SFH1-A, even if neither of them
is fully satisfactory.

\begin{figure*}[t]
\centering
\includegraphics[width=14cm]{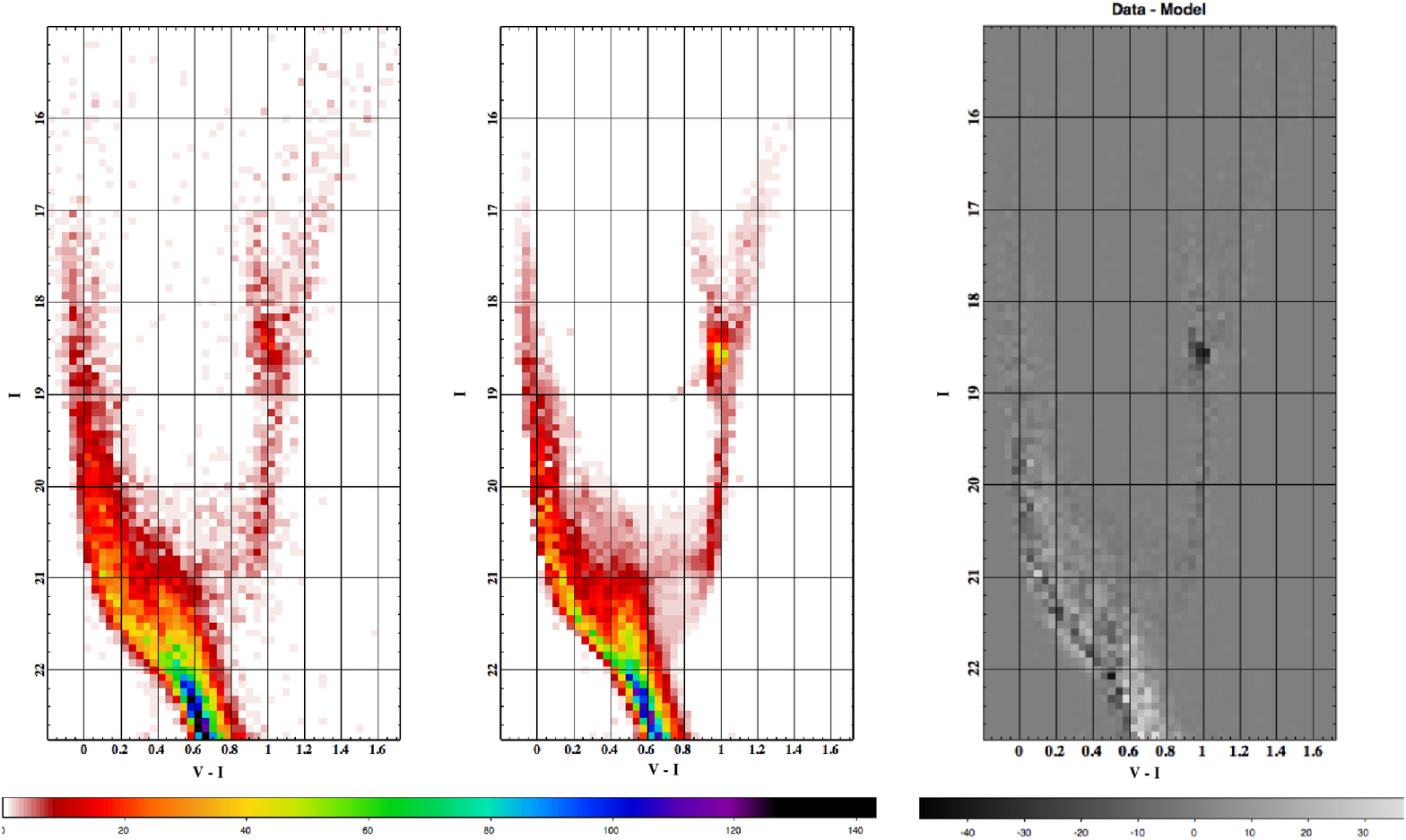}
\caption{Left panel: observed Hess diagram for the SFH1 field of the
  SMC Bar, created by binning the data by 0.04 in color and 0.08 in
  magnitude; middle panel: synthetic CMD from the SFH1-C model (see
  text for details); right panel: difference Hess diagram between the
  data and SFH1-C solution.}
\label{fig-colecmd1}
\end{figure*}

The result of applying Cole's CMD-fitting procedure to SFH1 is shown
in Figure~\ref{fig-colecmd1}.
   The left panel shows the data, binned
0.04$\times$0.08 in (m$_{V}-$m$_{I}$, m$_{I}$), while the middle and
the right panels show the best-fit synthetic CMD (SFH1-C) and map of
residuals, respectively. The fit procedure has matched the LF and
mean color of the stellar sequences well in general.  The data and the
model can be readily distinguished from one another because the data
is just one instance of a random draw from the parent population and
contains unmodelled noise, while the model represents the best guess
at the pure parent population and is therefore more smoothly
distributed.

Among the well-matched features are the color and vertical extent of
the RC and its upward extension to I $\approx$17, the enhancement in
the SGB at I $\approx$20.8, and the nearly-vertical finger of stars at
V$-$I $\approx$0.5 corresponding to an intermediate-age MSTO.  Notable
areas of mismatch include the failure to reproduce the width of the MS
(most obvious for I$\lesssim$21), and the factor of 2 overproduction
of RC stars (I $\approx$18.5, V$-$I $\approx$1).  There is a sparsely
populated red HB in the model that is not apparent in the data; this
may be related to the general factor of 2 overproduction of low-mass
core helium-burning stars, to poor constraints on the detailed element
abundances at ancient times, and/or to gaps in modeling the physics of
HB envelopes.

\begin{figure*}[t]
\centering
\includegraphics[width=9cm]{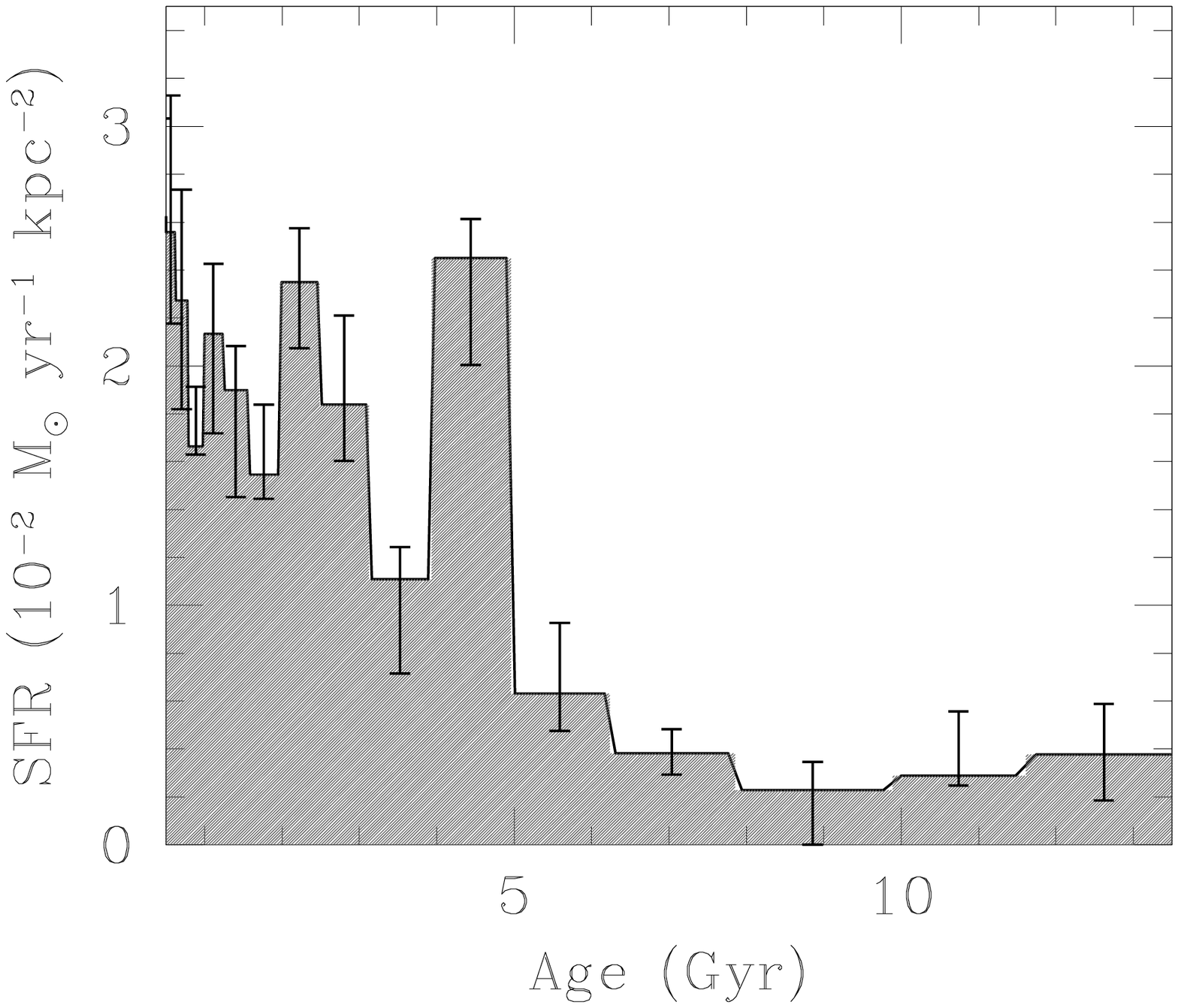}
\caption{Long term history of the SFH1 field from solution SFH1-C.
  Significant star formation commenced $\approx$5~Gyr ago and the rate
  has remained high, with some fluctuation, ever since.}
\label{fig-colesfh1}
\end{figure*}

The SFH over the period from 0.5 to 13.5 Gyr ago for SFH1-C is given
in Figure~\ref{fig-colesfh1}.  The SFH is characterized by a very low
star formation rate for several billion years before a rapid increase
about 5~Gyr ago.  This is the event that produces the features in the
MS and SGB of the SFH1 CMD.  The SFH remains at a similar elevated
level after the 5~Gyr event, with some fluctuations.

Overall, the SFH of the SFH1 field, based on the synthetic CMDs from
the Bologna and Cole procedures, is characterized by the following
features:

\begin{itemize}

\item The first 6--8 Gyr were rather quiescent and only a small
  fraction of the stars in this field was formed during these old
  epochs.  The inferred average rate of star formation for ages older
  than $\approx$5~Gyr is only $\approx$3
  $\times$10$^{-3}$~M$_{\odot}$~yr$^{-1}$~kpc$^{-2}$.

\item About 5--6 Gyr ago SFH1 experienced a remarkable enhancement
  in the stellar production: over $\approx$1~Gyr the activity ramped up
  to about $2.2 \times$ 10$^{-2}\,$M$_{\odot}\,$yr$^{-1}\,$kpc$^{-2}$.  The
  age and magnitude of this enhancement is robust to choices of IMF,
  CMD gridding, reddening, assumed metallicity, and details of the 
  fitting procedure.  


\item The average SFH has not dropped significantly from its peak at
  5~Gyr ago, but the degree of burstiness in the solutions is
  model-dependent.  SFH1-B shows a relatively smooth recent history,
  while SFH1-C shows a factor of two drop in SFR between 3--4~Gyr ago
  with a subsequent recovery and SFH1-A shows repeated gasps from
  0--2~Gyr ago.  The reasons for this range of behavior are considered
  in section~\ref{sec-compare} below.

\end{itemize}

\begin{figure*}[t]
\centering
 \includegraphics[width=9cm]{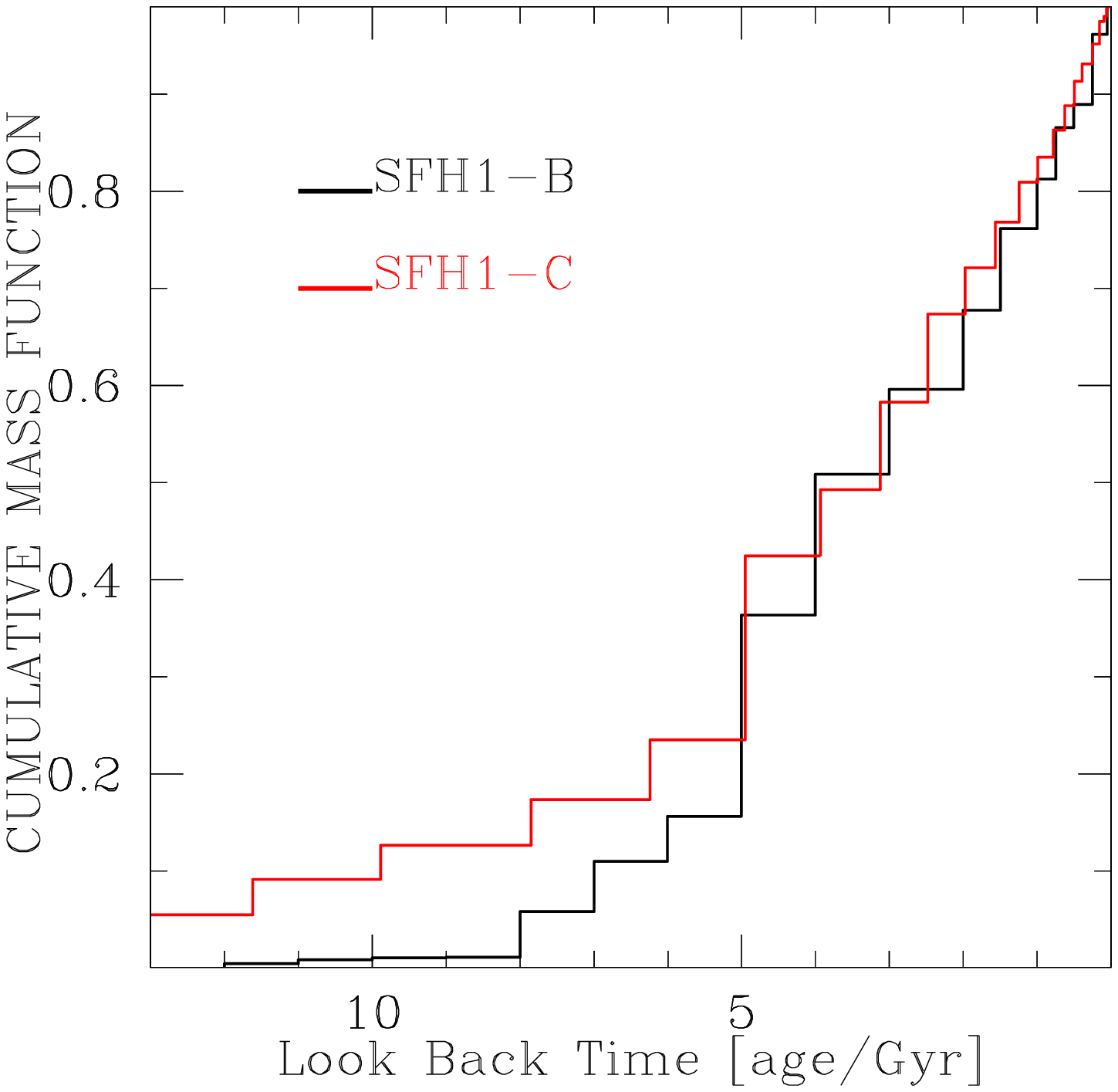}
\caption{Cumulative mass function in SFH1 according to the solutions
  SFH1-B and SFH1-C.}
\label{massa} 
\end{figure*}

It is worthwhile to quantify the stellar mass produced in each age
interval (see the cumulative mass function in Fig.
\ref{massa}). According to solution SFH1-B, the SFH1 field assembled a
small fraction (11\%) of its stellar mass before 6 Gyr ago, while a
significant fraction (about 40\%) was assembled only in the last 2
Gyr.  For comparison, SFH1-C predicts that 17\% of the stars were
formed prior to 6~Gyr, and 33\% over the past 2~Gyr.


\subsection{SFH4 field}

\begin{figure*}[t]
\centering
 \includegraphics[width=9cm]{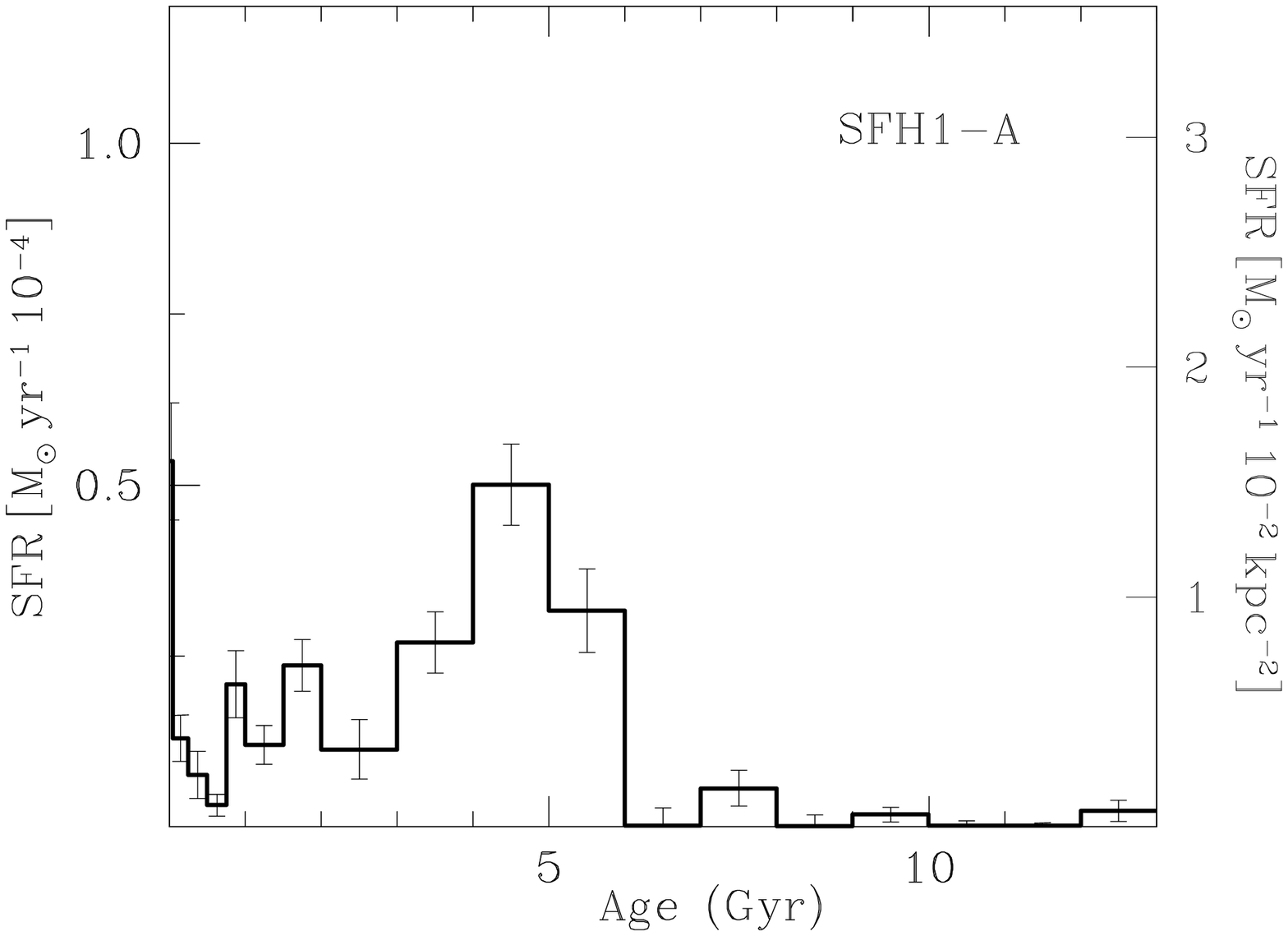}
\caption{Best recovered SFH with restricted metallicity range
(solution SFH4-A) for the field SFH4.  }
\label{sfr_f4} 
\end{figure*}

The history of the SFH4 field was determined in similar fashion to the
SFH1 field, using three sets of simulations (see
Table~\ref{tab-models}).  The recovered SFH for case SFH4-A is shown
in Fig. \ref{sfr_f4}.

\begin{figure*}[t]
\centering
 \includegraphics[width=12cm]{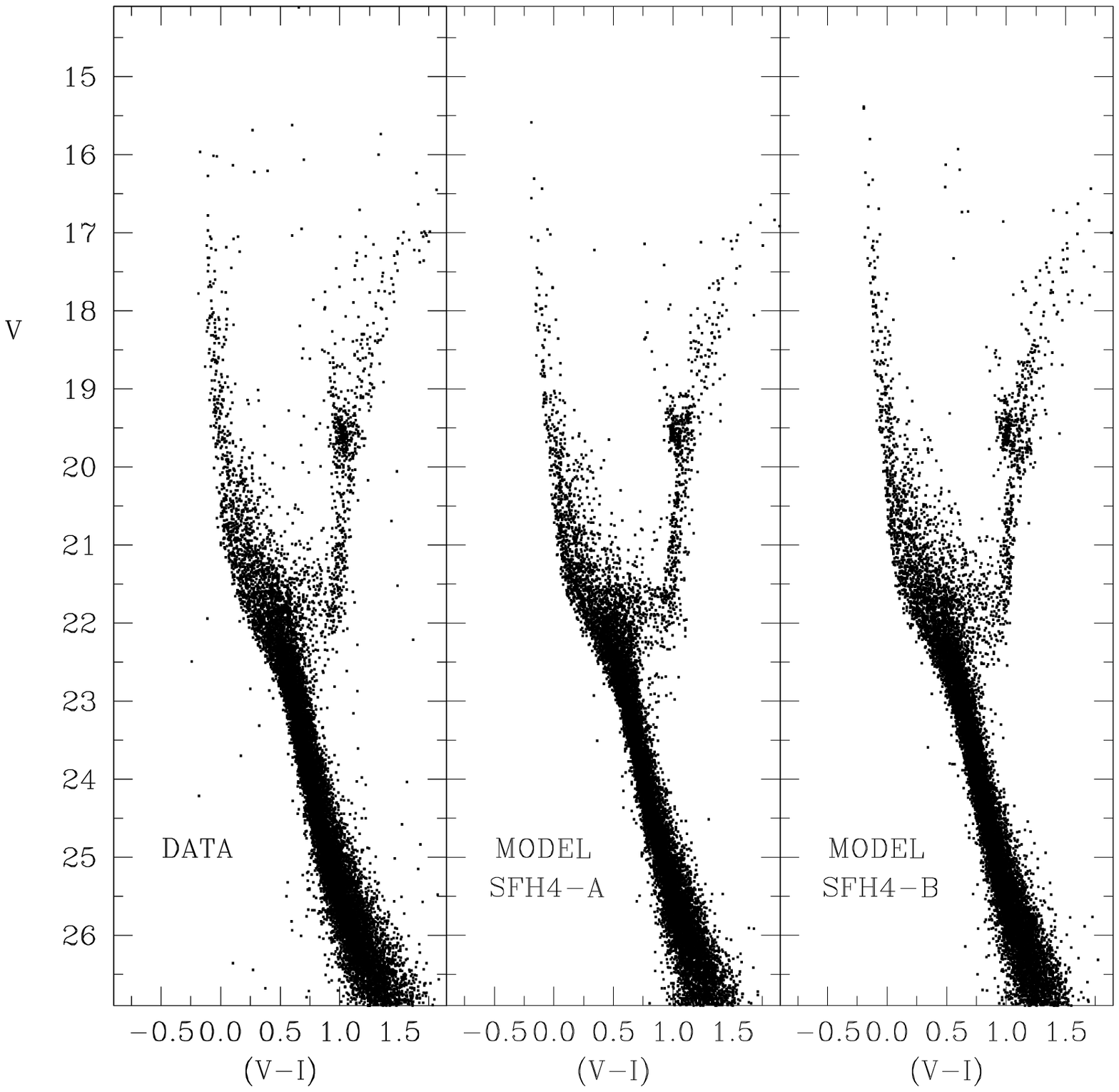}
\caption{Synthetic CMD for the solution SFH4-A (middle panel) and
  SFH4-B (right panel) compared with the observational CMD (left
  panel).}
\label{cmd_a_b} 
\end{figure*}

As in SFH1, the star formation proceeded at a low level until 5--6 Gyr
ago, when it rose to almost the same amplitude of the first peak in
SFH1. Afterward, the actvity started a slow but steady decline until
now and only in the last 50 Myr it reached higher levels (mimicking in
this respect the behavior of SFH1-A). Concerning distance and
reddening, the recovered values ($(m-M)_0=18.80$ and $E(B-V)=0.11$)
are not significantly different from those obtained for SFH1. The
corresponding reduced $\chi^2$, about 1.8, is significantly lower than
the best value obtained for SFH1. Indeed, a visual inspection of the
synthetic CMD (middle panel of Fig. \ref{cmd_a_b})
confirms a very good agreement with the observational counterpart: our
best model can reproduce the position and the morphology of the MS,
the SGB and the RGB (above and below the RC). The only minor
mismatches are in the BL region, which is underpopulated in our model,
and in the SGB luminosity distribution, which is slightly more
discontinuous than in the data.

\begin{figure*}[]
\centering
 \includegraphics[width=16cm]{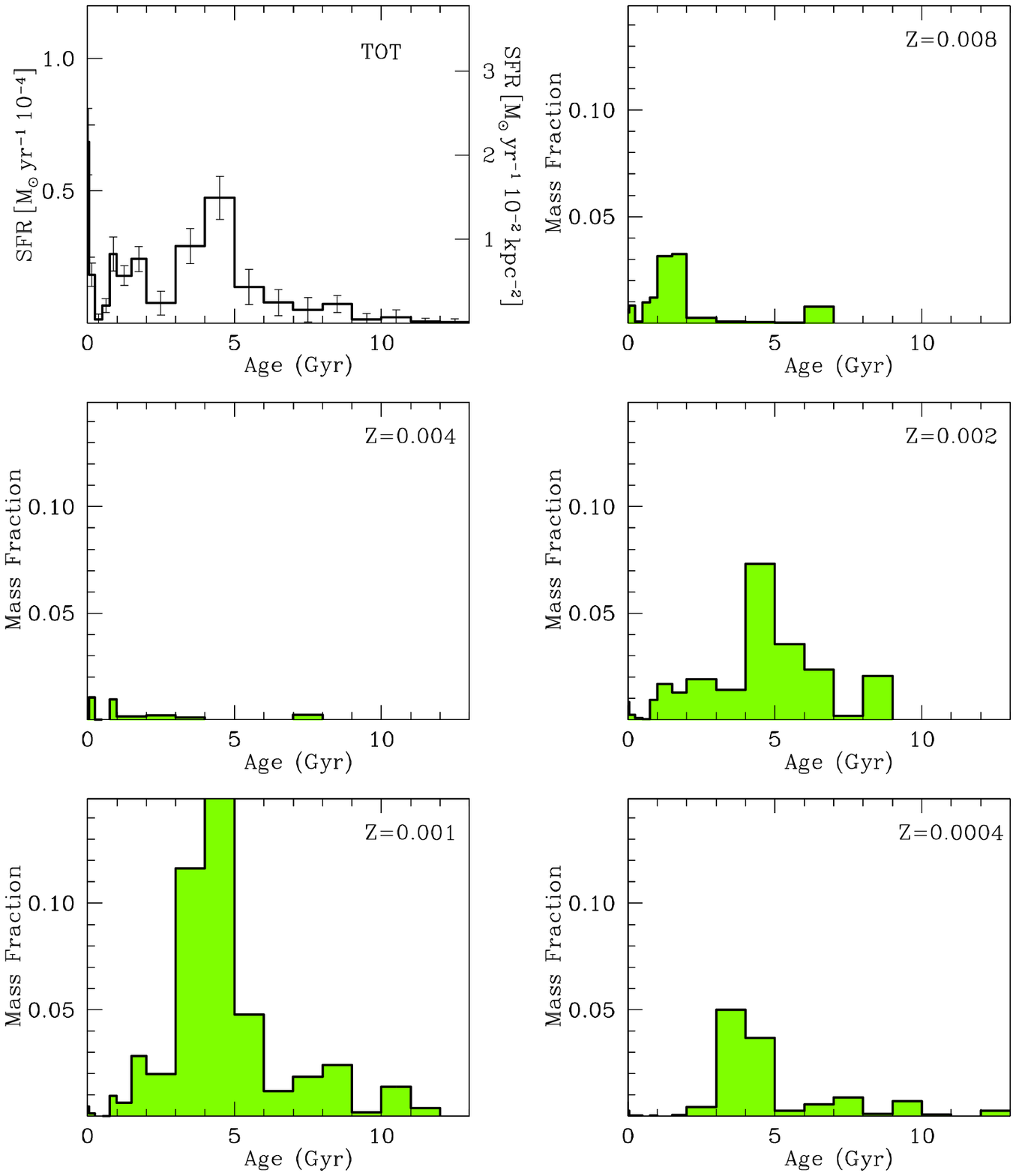}
\caption{Best SFH (solution SFH4-B) resulting from models which
  include also the metallicities $Z=0.008$ and $Z=0.0004$. The
  top-left panel shows the total SFH while the others display the
  contribution from each of the labeled metallicities.}
\label{sfr_f4B} 
\end{figure*}

\begin{figure*}[t]
\centering
\includegraphics[width=9cm]{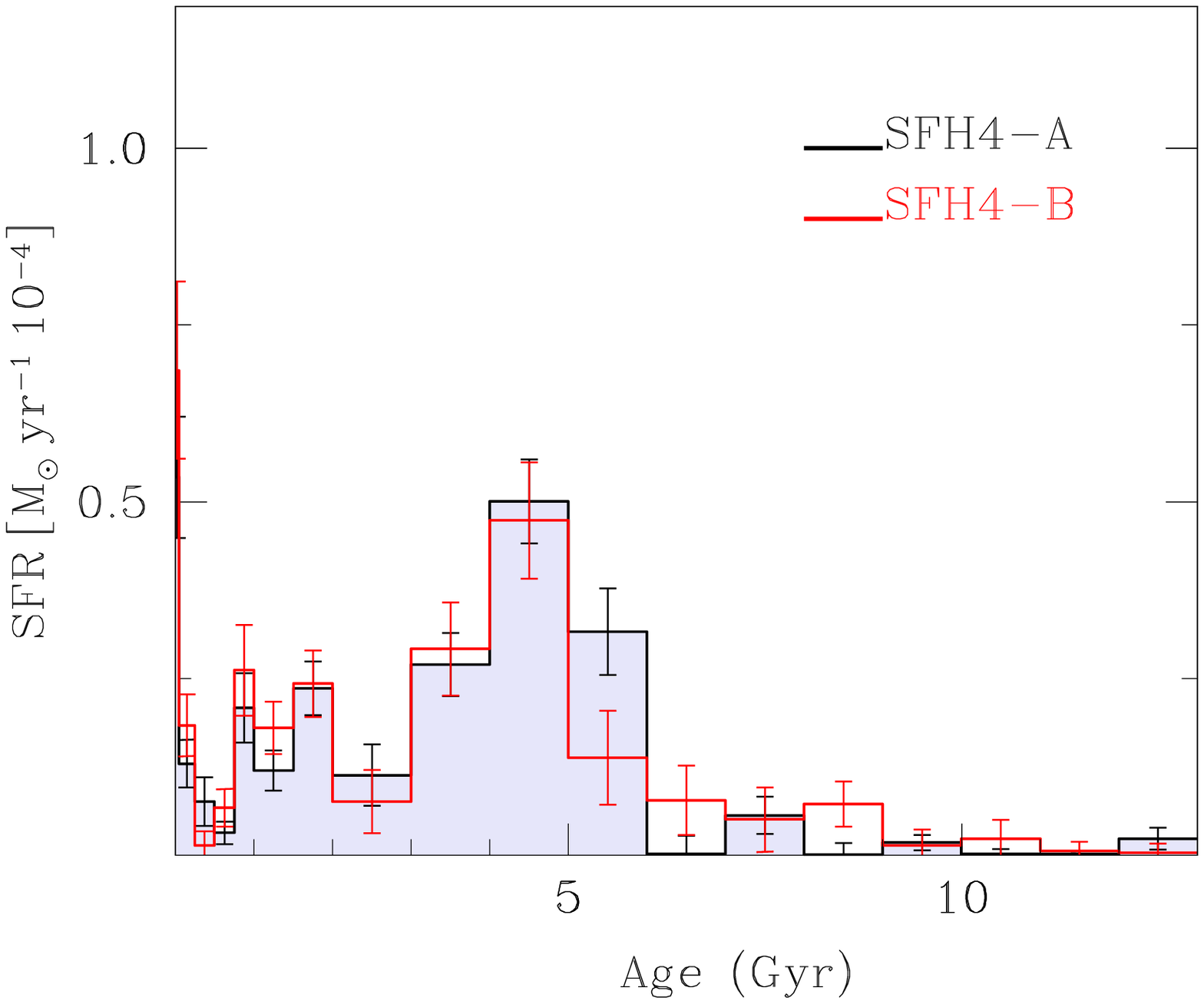}
\caption{Comparison between the solutions SFH4-A and SFH4-B.}
\label{sfr_f4a_f4b} 
\end{figure*}

As for SFH1, we expanded the metallicity range to include the values
$Z=0.008$ and $Z=0.0004$ and re-derived the SFH. The essential
features of the new solution (SFH4-B; see Fig. \ref{sfr_f4B}) 
are not changed (see Figure \ref{sfr_f4a_f4b} for a
comparison SFH4-A vs SFH4-B): there is still a long quiet initial
period, followed by a rapid enhancement of the activity and a
subsequent decline. According to the solution SFH4-B, most of the
stars ever formed in SFH4 have metallicities around $Z=0.001$.

\begin{figure*}[t]
\centering
 \includegraphics[width=10cm]{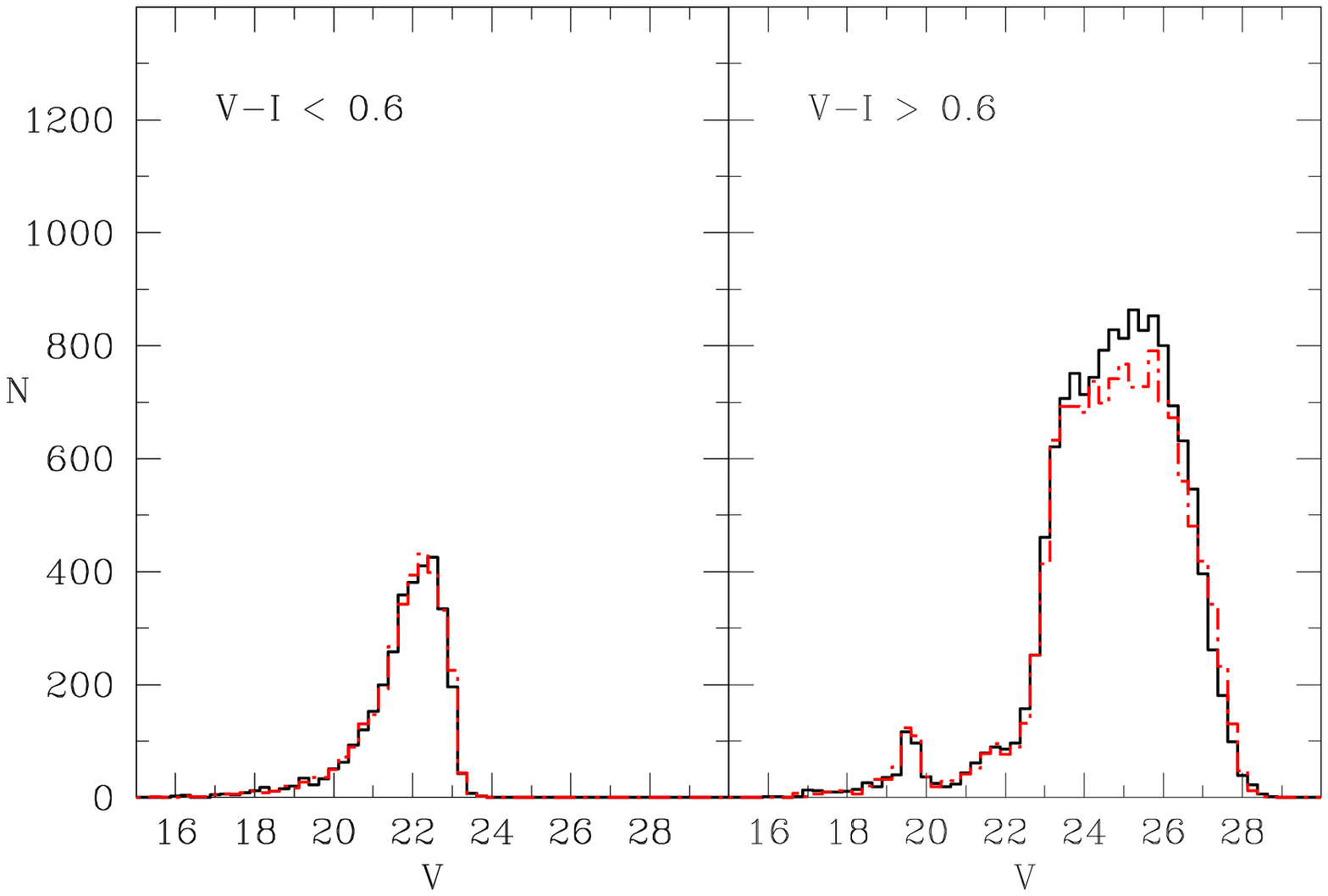}
\caption{Predicted LF (red histograms) for the solution SFH4-B vs the
  observed (black histograms) LF. The left-hand and the
  right-hand histograms are for stars bluer and redder than $V-I=0.6$
  respectively.}
\label{LF_f4} 
\end{figure*}

\begin{figure*}[t]
\centering
\includegraphics[width=16cm]{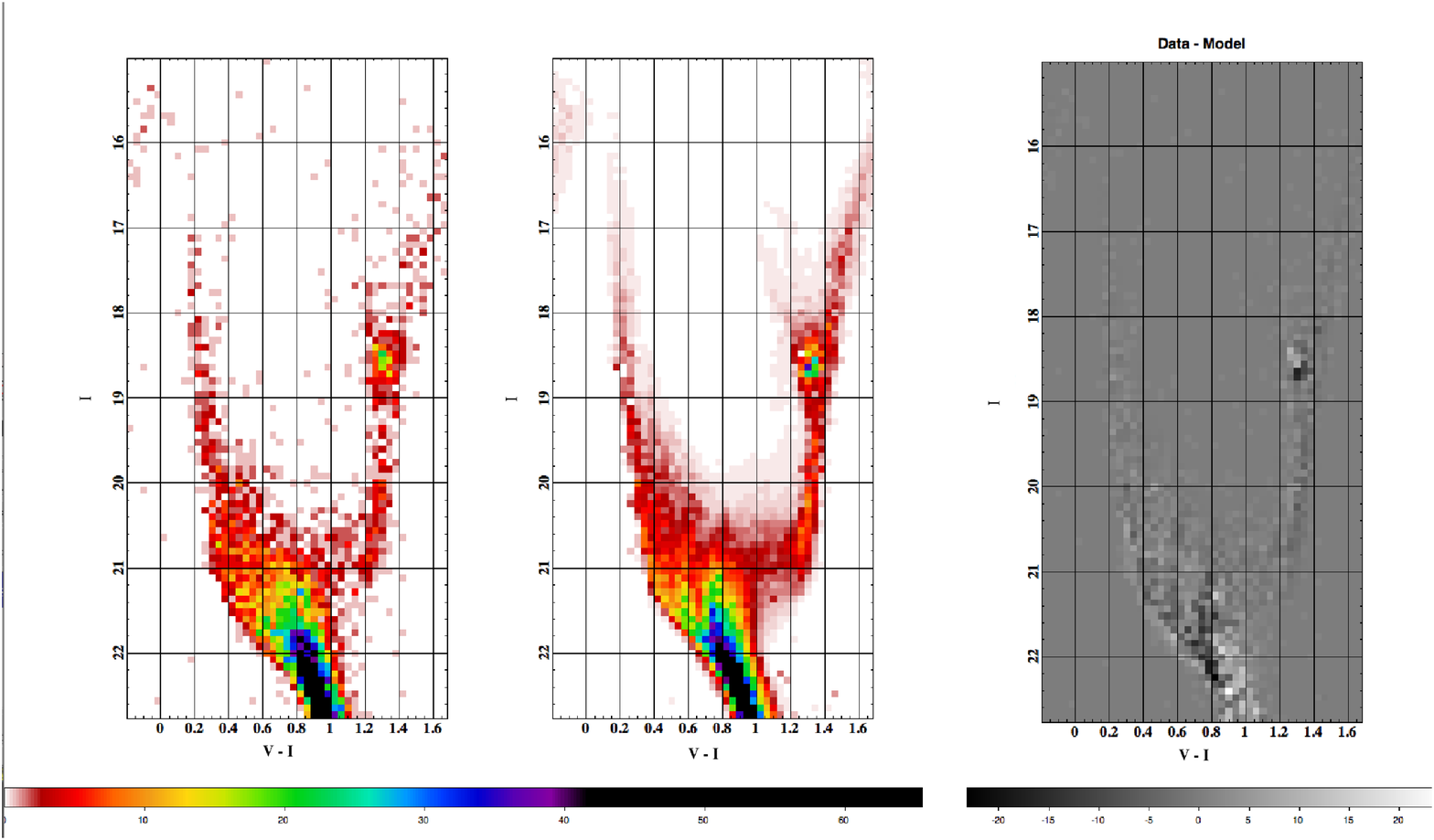}
\caption{Left panel: observed Hess diagram for the SFH4 field; middle
  panel: synthetic Hess diagram from solution SFH4-C; right panel:
  difference between the two.}
\label{fig-sfh4hess}
\end{figure*}

In terms of fit quality, solution SFH4-B improves the match to the
data significantly, leading to a reduced $\chi^2$ of about 1.3.  The
corresponding synthetic CMD is excellent (Fig.\ref{cmd_a_b}, right
panel) and the only detectable discrepancy is around the BL phase,
still slightly underpopulated by the model. Unlike in SFH1, the number
of RC stars is well matched (see the LF in the right panel of
Fig. \ref{LF_f4}). 
The fit to SFH4 using Cole's method shows similar
results, with a better fit to the upper MS than in SFH1 because the
observed sequence is narrower in color, and an overproduction of RC
stars.  The observed Hess diagram, synthetic diagram from solution
SFH4-C, and the difference between the two are shown in
Figure~\ref{fig-sfh4hess}.

\begin{figure*}[t]
\centering
\includegraphics[width=9cm]{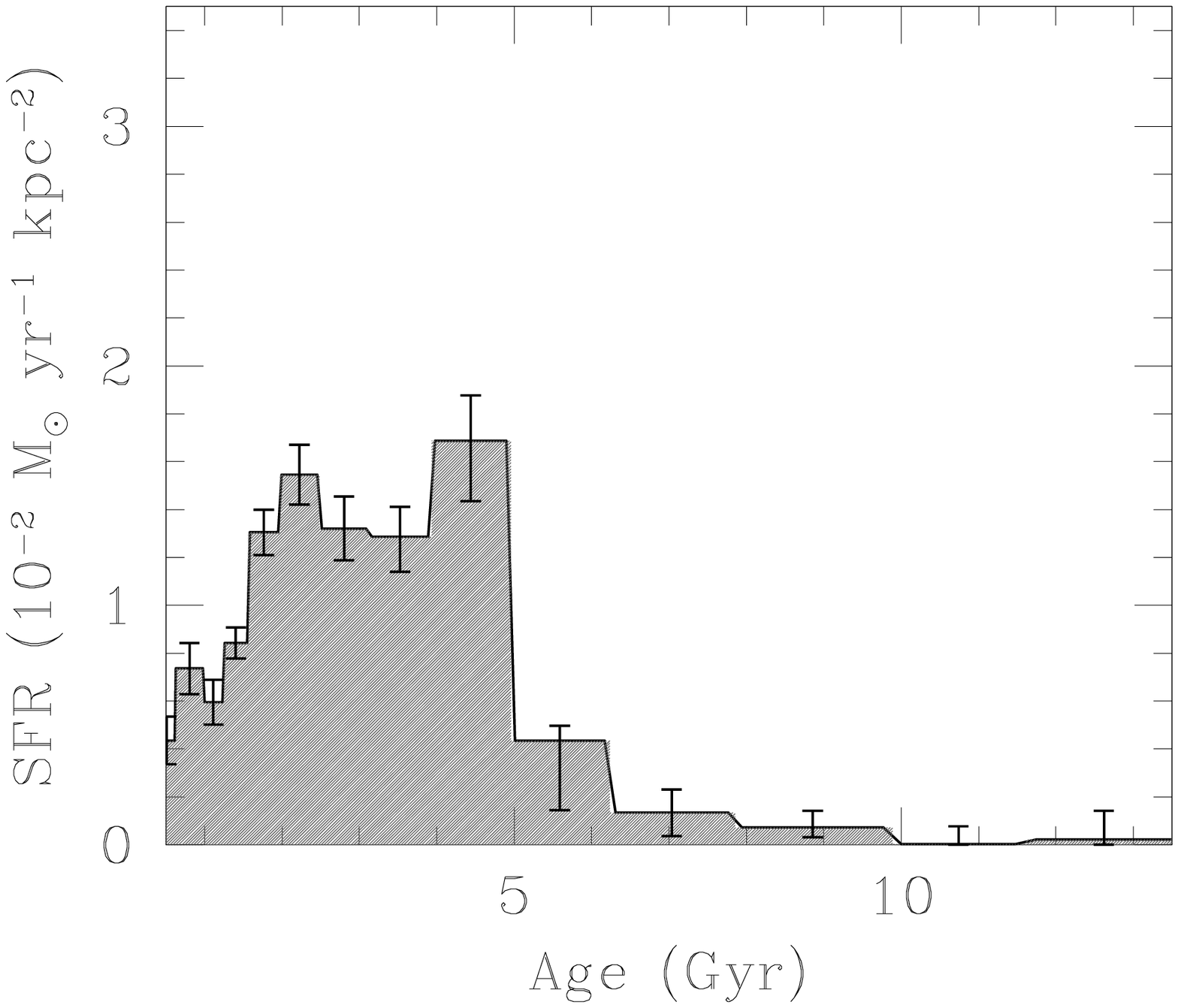}
\caption{Long term history of the SFH4 field from solution
  SFH4-C. Similar to SFH1, the field ``switched on'' $\approx$5~Gyr
  ago. The SFR has been declining over the past $\approx$1.5~Gyr in
  this locale. } 
\label{fig-colesfh4}
\end{figure*}

SFH4-C is broadly similar to SFH1-C, but the mean SFR is lower at all
ages, commensurate with the smaller number of stars in the field
(Figure~\ref{fig-colesfh4}). 
It is notable that the onset of significant star formation occurs
simultaneously in the two fields, to within the $\pm$1~Gyr precision
of our data.  However, the SFH4 field shows a decrease in SFR relative
to SFH1 over the last $\approx$1.5~Gyr (which is obvious from a
comparison of the RC
morphology and upper MS LF of the two fields).\\

\begin{figure*}[t]
\centering
 \includegraphics[width=9cm]{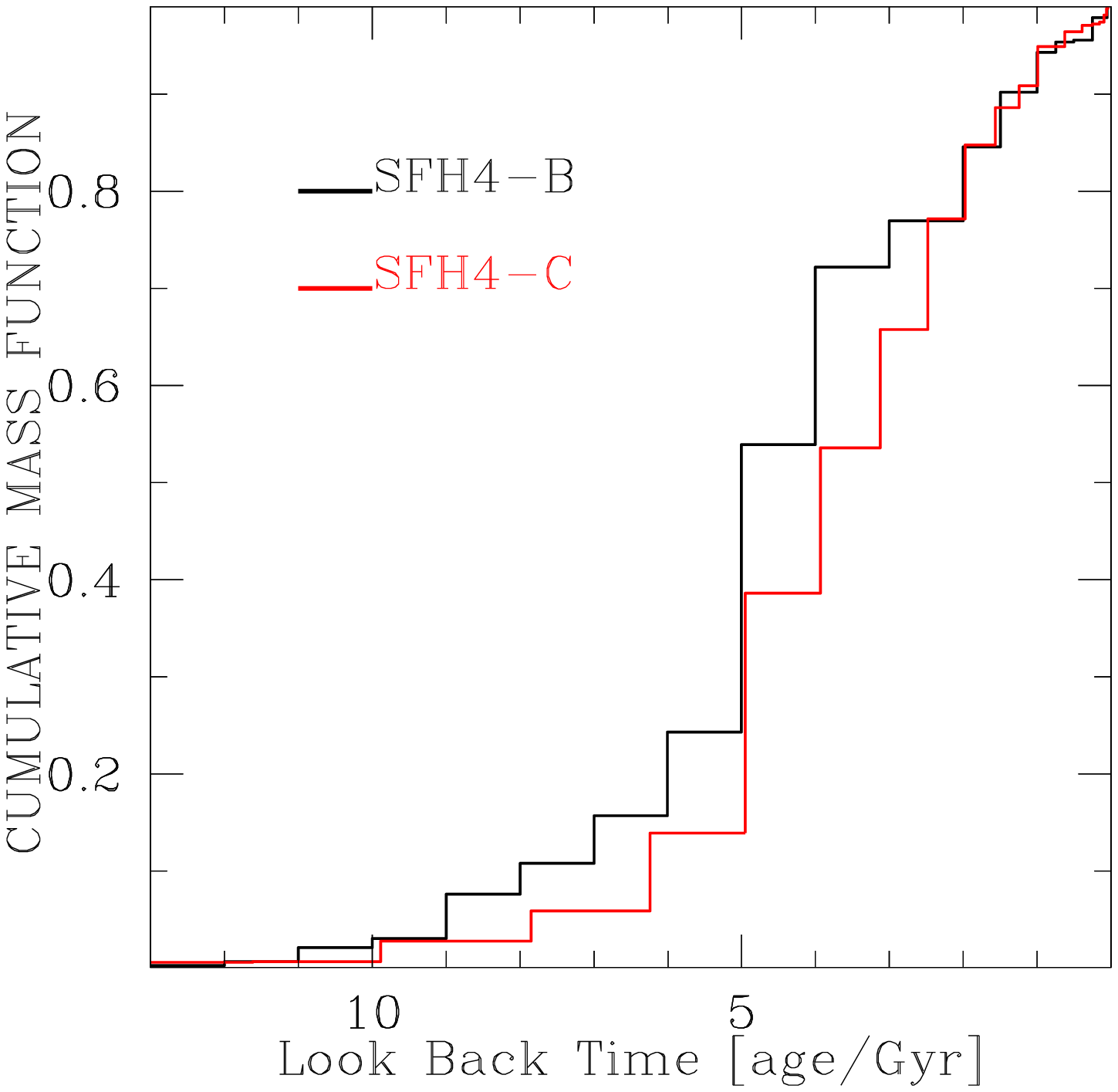}
\caption{Cumulative mass function in SFH4 according to the solutions
  SFH4-B and SFH4-C.}
\label{massa_f4} 
\end{figure*}

According to the solution SFH4-B (see the cumulative mass function of
Figure \ref{massa_f4}) 
about 60\% of the stellar mass has been produced between 6 Gyr and 2
Gyr ago, which is higher than the fraction of mass (48\%) produced in
SFH1 in the same period. On the other hand, 16\% of the total mass was
already in place before 6 Gyr ago, which is comparable with the
production in SFH1 (11\%). The former difference is compensated in the
last two Gyr, when SFH1 has turned into stars about twice the mass
fraction of SFH4.  While in the SFH1 field the Bologna solution
produced a younger result than the Cole solution, in SFH4 the
situation is reversed.  The solution SFH4-C predicts that only 7\% of
the stellar mass was in place before 6~Gyr ago, while 29\% was
astrated in the past 2~Gyr.

\subsection{Comparison of the two methods \label{sec-compare}}

\begin{figure*}[t]
\centering
\includegraphics[width=17cm]{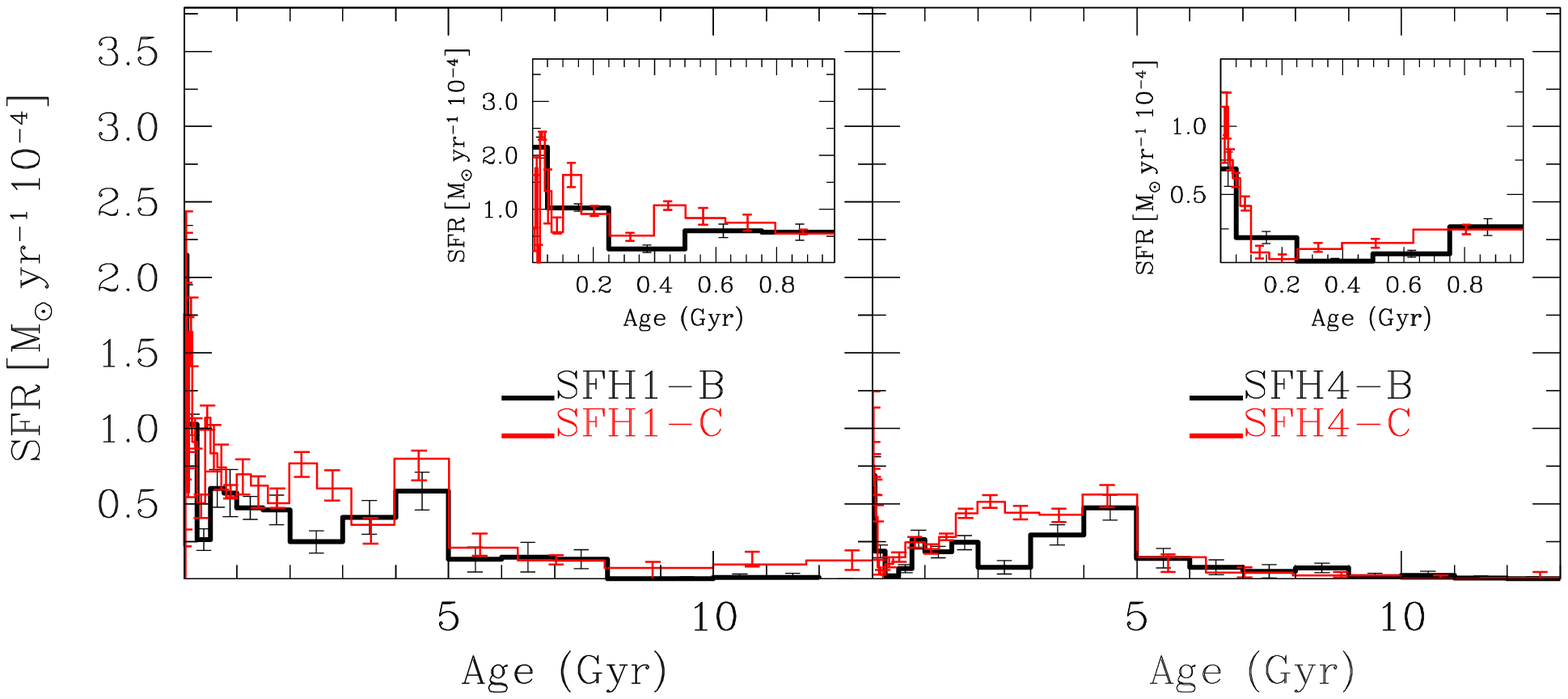}
\caption{Bologna SFHs (black line) compared to Cole's (red
  line). The left panel and the right panel refer to the fields SFH1
  and SFH4, respectively. The inset panels zoom in on the last 1 Gyr.}
\label{bo_cole} 
\end{figure*}

We have derived the SFH of the field stars in the SMC Bar using two
different synthetic CMD techniques. Figure~\ref{bo_cole} 
shows the results plotted together on the same scale to illustrate the
similarities and differences. For SFH1 both SFHs indicate a major
activity in the last 5 Gyr, an overall plateau (on average) since then
on, and a spike in the last 50 Myr. The major differences concern: 1)
The exact epochs of the peaks between 250 Myr and 4 Gyr ago (around
1.5 Gyr ago in Bologna's solution and 2.5 Gyr ago in Cole's solution);
2) The rate in the period 8-13 Gyr ago, which is slightly stronger in
Cole's solution.

The agreement for the field SFH4 is good as well: both models predict
a star formation onset between 4 and 5 Gyr ago, followed by a slow
decline and a very recent burst. The major difference here is the rate
of decline in SFR after the 4--5~Gyr enhancement, with Cole's solution
showing a constant activity for about 2.5 Gyr and Bologna's solution
showing a faster decline followed by a mild enhancement around 1.5 Gyr
ago.

In both fields, the only ages in which the star formation rates differ
by significantly more than the formal errorbars on the solutions are
in the period 1.5--3~Gyr ago, with Cole's solution showing
consistently higher SFR. Part of this difference stems from the
slightly higher SFR derived by Cole's method at all ages, owing to the
different mean stellar masses resulting from the different assumed
IMFs and binary fractions.  The effect may be exacerbated in the
1.5--3~Gyr age range because of the interplay between age and the
different metallicity values considered, and the fact that Cole's
procedure does not consider the RC in calculating the best-fit SFH.
Consideration of a very large number of metallicity values tends to
produce smoother SFHs than those derived using only a few
metallicities \citep[e.g.,][]{cole09}.

Concerning the distance modulus, Cole's best values (18.83 for SFH1
and 18.85 for SFH4) are slightly higher than Bologna's (18.77 and
18.80).  In SFH1 these offsets may be due to the higher reddening
required by the Bologna method, but in SFH4 the converse is true. Yet
the distance modulus offsets are similar.  The offsets are quite small
and can be considered to be within the noise, but a possible reason
could be the fact that Cole's method ignores the RGB and RC, so the
distance is essentially a MS-fitting distance, while the Bologna
method includes RGB and RC information.  More interestingly, both
distance estimates are shorter than recent determinations based on
RR-Lyrae ($(m-M)_0=18.90$, \citealt{kapakos2011}) and eclipsing
binaries ($(m-M)_0=19.11$, \citealt{north2010}), but still compatible
with the average distance of star clusters (around 18.87 for
\citealt{glatt08b} and between 18.71 and 18.82 for
\citealt{crowl2001}). This variance may indicate a different
sensitivity to reddening, which is highly variable across the SMC
\citep[see e.g.][]{zaritsky2002, haschke11}, or a line of sight depth effect (up
to 4.9 kpc according to \citealt{subra2009}, between 10 and 17 kpc
according to \citealt{glatt08b}) which may depend on the direction.

\begin{figure*}[t]
\centering
\includegraphics[width=10cm]{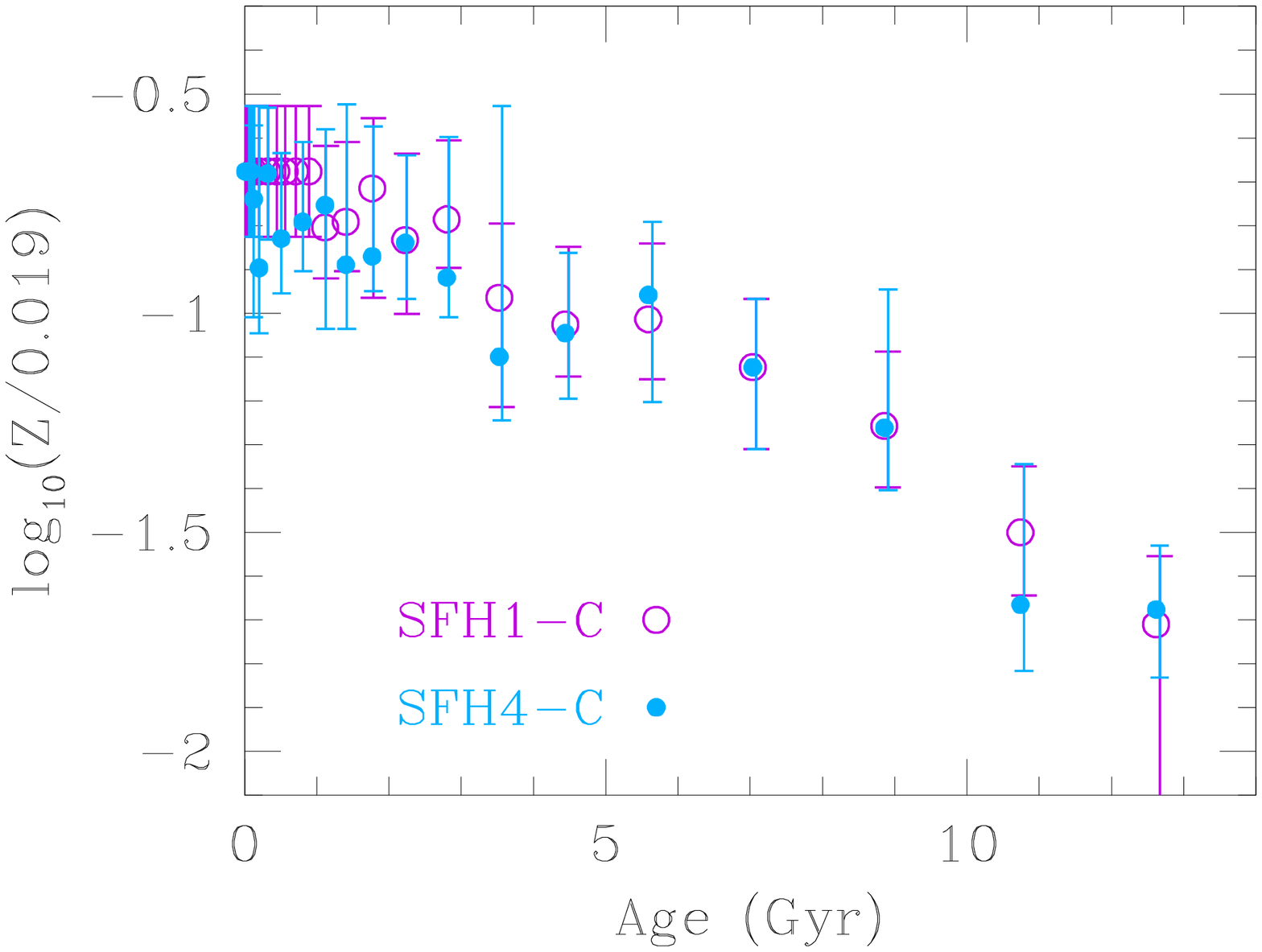}
\caption{Best-fit age-metallicity relation of the two Bar fields derived
by Cole's method.  The two fields show similar chemical evolution; within
the errorbars they are the same.  Taken together they paint a consistent
picture of average metallicity increasing steadily over time.}
\label{fig-coleamr}
\end{figure*}

The age-metallicity relations inferred by both methods are consistent,
implying a metallicity slowly increasing with time and the bulk of the
stars with metallicities between 0.001 $\lesssim$ Z $\lesssim$ 0.002.
The models derived with the Bologna procedure consistently find better
fits when a metallicity of Z = 0.008 is included for the younger ages,
while such a population is not required in Cole's models (see
Fig.~\ref{fig-coleamr}).  
However, Cole adopted a higher differential
reddening in his modeling of SFH1, which has the higher fraction of
young stars; see Table~\ref{tab-models}.

\subsection{Comparison Between the Two Fields}

\subsubsection{Spatially Resolved SFH of the SMC Bar}

Direct comparison of the two fields shows that colors and magnitudes
of the prominent stellar sequences are similar, and the main
differences concern the younger stellar populations.  In SFH1 the
upper MS, BL, and red supergiant region are more populous, indicating
a higher SFR at ages of a few hundred Myr or less.  A challenge for
both SFH-fitting methods was to reproduce the breadth of the upper MS
in SFH1, as well as its RC, which was overproduced by both procedures.
Integrating up the predicted stellar masses from the SFH-fitting
procedures we find that $\approx$4$\times$10$^5$ M$_{\odot}$ stars
formed over the lifetime of SFH1, and 2$\times$10$^5$ M$_{\odot}$ in
SFH4.

The general similarity of the two CMDs indicates that their features
are representative of the general properties of the central SMC over
most of its history. Small localized bursts/gasps of star formation
may be smoothed away by the drifting of stars away from their
birthplace except for the youngest ages. The dramatic increase in SFR
at 5~Gyr is a strong feature in both fields and suggests a global
change in environment to one that strongly favored star formation
throughout the SMC at this time.  Backward integration of the
Magellanic Clouds orbital paths around each other and the Milky Way
over this length of time are quite uncertain, and it is therefore
difficult to say if this increase is associated with a tidal
interaction or not.

More recent dynamical signatures should be easier to locate in time,
and thus to correlate with global star formation
events. \cite{diaz2011} presented orbital calculations for the system
MW/LMC/SMC, finding evidence for a close encounter about 1.5-2.0 Gyr
ago. Analyzing SMC clusters, \cite{piatti2011} also suggested two
enhanced formation processes that peaked at 2 and 5-6 Gyr ago. A
comparison with our solutions suggests that the former episode might
be associated with the Bologna (1.5 Gyr) and Cole (2.5 Gyr)
intermediate age peak, while the latter coincides with our strong rise
5 Gyr ago.

The inset panels in Fig. \ref{bo_cole} show the recent 1~Gyr history
of both fields according to Bologna and Cole methods. In the SFH1
field the mean specific SFR over the past 1~Gyr is nearly as high as
the Gyr-averaged peak rate during the 5~Gyr event, but in SFH4 the
recent SFR is reduced by a factor of $\approx$3 from the
intermediate-age peak.  However, in both fields the recent average
specific SFR is still several times higher than the rate prior to the
major episode 5~Gyr ago.  The increase in the very youngest age bin
may be an artifact of the fitting process, because the only stars
younger than $\approx$50~Myr in our CMDs are still on the MS - all of
the evolved stars are above our bright limit- and so the code is free
to vary the SFR at young ages somewhat arbitrarily to fit the
residuals left after older populations are constrained. However, in
literature there are also reports indicating very recent bursts of
activity; see e.g. the analysis by \cite{indu2011} who found peaks at
0-10 Myr and 50-60 Myr ago. Stars formed in these events, however, may
not have had time to diffuse through the SMC, and thus may not be
present in our two small HST fields.

There are also clear differences between the two fields, with SFH4
showing a lower overall SFR density averaging
$\approx$5$\times$10$^{-3}$ M$_{\odot}$~yr$^{-1}$~kpc$^{-2}$ and SFH1
forming stars at the average rate of 2.8$\times$10$^{-2}$
M$_{\odot}$~yr$^{-1}$~kpc$^{-2}$ with an apparently higher degree of
burstiness.  The features in the CMD that indicate high SFR at the
``burst'' ages are the large number of MS stars at I $\approx$17--17.5
(100--150~Myr) and the steepening of the MS LF at around I
$\approx$19.5--20, with the resulting large number of red core
helium-burning stars (vertical RC; 400--800~Myr).  In SFH1 the average
recent SFR is comparable to the peak of the longterm average SFH since
the 5~Gyr episode, but in SFH4 the rate has dropped by a factor of
$\approx$3 from the peak.  If it is proposed that a tidal interaction
between the SMC and LMC (e.g. \citealt{besla2012}) or Milky Way
resulted in a period of increased SFR at 100--200~Myr as seen in SFH1,
then it remains to be explained why this event is not seen in the SFH4
field.

\subsubsection{Chemical Evolution}

Because a range of metallicities is considered at most ages, we can
plot the resulting age-metallicity relation for the best-fit
solutions.  This is shown for the Cole solutions (which use a finer
grid of metallicities, see Table~\ref{tab-models}) in
Figure~\ref{fig-coleamr}, which shows that the two fields have
experienced similar longterm chemical evolution histories.  This is
not a surprising result because over time periods longer than
$\sim$10$^{9}$ yr the entire galaxy should be mixed, but it is a
confirmation that the fit procedure is giving consistent results
despite the differences between the two fields. For the youngest
populations, only a single metallicity (Z=0.004) has been used, so the
errorbars simply reflect the spacing of the metallicity grid.  Where
multiple metallicities were used, the mass-averaged mean metallicity
is given, and the error bar shows the range of metallicities within
which 90\% of the stars in each age bin are expected to fall.  The
mean metallicity is quite similar to that obtained from the Bologna
method results, with the exceptions that Z=0.008 populations are not
present in large numbers in the Cole solution, and the Bologna
solution for SFH4 appears to increase its metallicity more quickly for
ages older than $\approx$6~Gyr.  Given the uncertainties in RGB model
colors and the differences in distance and reddening between the two
fits, it is not clear these differences are significant.

The field star AMR shown in Fig.~\ref{fig-coleamr} is consistent with
\citegenitive{piatti2012} AMR, whose AMR is obtained from CCD
Washington CT1 photometry.  Moreover the metallicity values for ages
older than 1~Gyr match well the spectroscopic field star measurements
of red giants in the central SMC reported by \citet{par10}.

  Compared to AMRs derived from clusters, our result is in good
  agreement with \citet{dac98} and \citet{glatt08b}, while it differs
  from the cluster AMR of \citet{piatti2011} and
  \citet{par09}. \citegenitive{piatti2011} metallicity dispersion is much
  higher than our formal errors at any age, suggesting that SMC
  clusters may have originated from a less mixed interstellar
  medium. \citegenitive{par09} AMR rises faster than ours prior to 10 Gyr,
  suggesting that SMC clusters may have experienced an independent
  chemical evolution history.

  It is worth to remind that we have derived the best AMRs from two
  small regions of the SMC and more fields are needed to confirm our
  results.  A detailed comparison of our six AMRs (sampling regions
of the Bar, Wing and Halo) with spectroscopic metallicity measurements
will be presented in a future paper.

\section{Comparison with other studies}

\begin{figure*}[t]
\includegraphics[width=17cm]{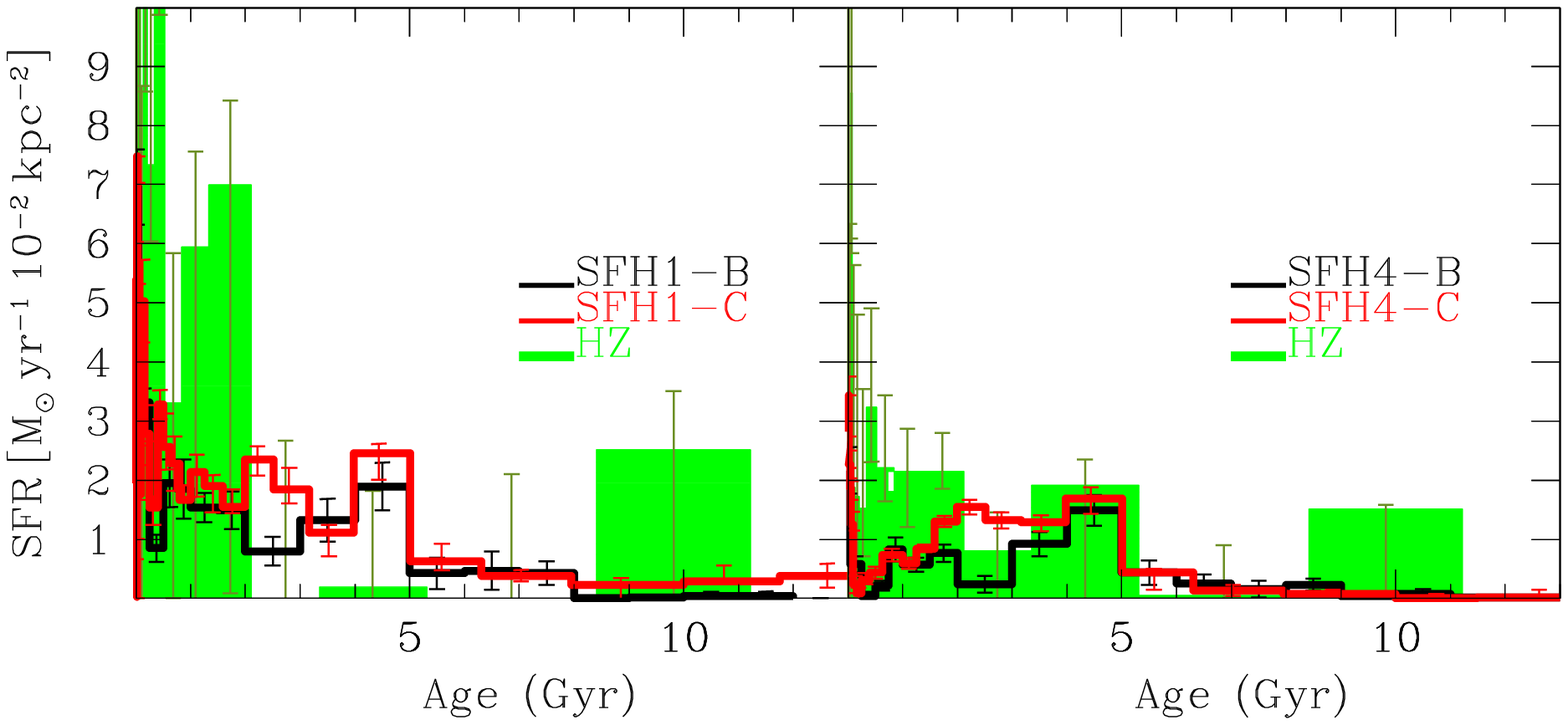}
\caption{Comparison between our results (Bologna's solution is in
  black, Cole's is in red) and \cite{harris04} solutions (HZ)
  for SFH1 (left panel) and SFH4 (right panel). }
\label{harris} 
\end{figure*}

\citet{harris04} (hereafter HZ04) were the first to apply the
synthetic CMD method to the derivation of the SMC SFH. They used
wide-field UBVI photometry of an area of $4\degr \times 4.5\degr$
(MCPS catalog, \citealt{zaritsky2002}), suggesting that 50\% of SMC
stars formed earlier than 8.4 Gyr ago, and that the period between 3
and 8.4 Gyr ago experienced very little, if any, star formation
activity. Figure \ref{harris} 
shows HZ04's results for their regions including our SFH1 and SFH4
compared with our derivations. Both HZ04's solutions show a
significantly higher stellar production prior to 8.4 Gyr ago. At
intermediate ages HZ04's solution for SFH1 is characterized by a clear
and long lull between 2 Gyr and 8.4 Gyr ago, which is in striking
contrast with our SFHs, whereas HZ04's SFH for SFH4 is stronger and in
satisfactory agreement with our predictions. In the last 2 Gyr both
HZ04's star formation rates are much stronger than our findings.

One should note that the HZ04 SFHs are derived from a much larger
field of view (about 16 times wider) and their photometry did not
reach the oldest MSTO. This last point is particularly relevant, since
studies that do reach it indicate, instead, that, although present,
stars older than 8 Gyr do not dominate the SMC population
\citep{dolphin01,mccumber05,noel07,noel09,chiosi07}. Further support
to this is also provided by the relatively low number of RR Lyrae
stars detected in the SMC compared to the LMC (\citealt{soszy10}).

\citet{dolphin01} analyzed an external region close to the globular
cluster NGC~121, and concluded that stars in the outskirts of the SMC
formed during a broadly peaked episode of star formation, with the
largest rate between 5 and 8 Gyr ago. As expected, this outer field is
significantly older than ours, although the contribution from the age
bins older than 10 Gyr is again a minor fraction of the total mass
produced.

\citet{mccumber05} studied an external field in the SMC Wing, and
found an increasing rate from 12 to 4 Gyr ago, and then over the past
1.5 Gyr, with a significantly quieter phase between 4 and 1.7 Gyr
ago. This analysis was not conducted using a statistical significance
testing and their favored solution was the one (among three ansatz
models) which best matched the star counts in five strategic CMD
regions. Although this makes the comparison with our finding
difficult, two considerations can be made: 1) also in this case the
stellar production prior to 10 Gyr ago is low, in agreement with our
result; 2) apart from a recent burst, the average activity of their
SFH is earlier than in our fields, as expected for a Wing region,
which represents a population with intermediate age between the Bar
and the outskirts.

\citet{noel07} and \citet{noel09} produced one of the most extensive
and accurate photometric campaign in the SMC outskirts (see also
\citealt{nidever2011}). Their 12 fields are sufficiently deep to reach
the old MSTO and none of them showed a clear extended HB,
representative of a very old and metal poor stellar
population. \citet{noel09} used a population synthesis technique and
recovered a great variety of SFHs, with only the outer fields
characterized by a strong old and intermediate age activity. Although
the spatial distribution of these fields, mainly enclosed in the
periphery of the SMC, makes the comparison with our results less
direct, we note that among \citegenitive{noel09} solutions, their
innermost field (qj0112) most closely resembles our history of the
Bar.  Prior to 8 Gyr ago, the typical SFR density in their fields
seems comparable with ours, suggesting that the SMC periphery and Bar
share a common old population but that the initiation of significant
star formation at 5--6 Gyr affected the Bar far more than the external
regions of the SMC, marking the birth of the Bar as a distinct
feature.

A more direct comparison can be made with \cite{chiosi07}, whose
analysis was focused on three deep HST/ACS fields located in the
Bar. In this case, our solutions are in good quantitative and
qualitative agreement. All \citegenitive{chiosi07} SFHs show both an
unambiguous rise between 7 and 5 Gyr ago (with the precise epoch
varying depending on the specific direction) and a negligible earlier
activity. However, despite the low rate, their and our early
  activities are not zero, offering a natural explanation to the
  relatively low number of RR-Lyrae stars found in the SMC. Moreover,
  no two solutions are alike, suggesting that the star formation rate
  at young and intermediate ages was strongly variable across the SMC
  Bar.

Our results are rather similar to the SFHs of other two Local Group
irregulars such as IC~1613 \citep{skillman03} and Leo A
\citep{cole07}: both galaxies show an initial quiet period followed by
a prompt rise in the SF activity about 5-7 Gyr ago, with a peak
between 3 and 6 Gyr and between 2 and 3 Gyr ago, respectively. In both
cases the recent activity ($<$ 2 Gyr ago) is more similar to that in
SFH4 than in SFH1, presumably because the IC~1613 explored field is in
the outskirt and the Leo A field is large enough to include central as
well as peripheral regions.

Finally, comparing our SFH with the solutions found in the literature
for the LMC (see e.g. \citealt{harris09} and \citealt{smecker2002}) we
find an overall similarity. Apart from the very early SF enhancement
predicted by both Harris et al. and Smecker-Hane et al. that has no
counterpart in our SMC solutions, our results for SFH1 share key
properties with \citegenitive{smecker2002} findings for the LMC Bar: 1) a
lull over the period 6-10 Gyr ago; 2) a steep increase at 5-6 Gyr; 3)
a gasping regime over the last 5 Gyr.

In spite of these similarities, the SMC and LMC AMRs (see
e.g. \citealt{carrera2008}, \citealt{harris09}) remain rather
different. While the LMC metallicity has increased faster both prior
to 10 Gyr ago and over the last 4 Gyrs (reaching $[Fe/H]\sim -0.7$ and
$[Fe/H] =-0.2$, respectively), between these periods it has progressed
at much slower rate.

  Moreover, it is well known that the LMC cluster distribution shows a
  long gap between 3 and 12 Gyr. If this gap is a consequence of the
  quiescent period between 6 and 10 Gyr ago (see
  e.g. \citealt{harris09}), then it is intriguing that a similarly
  long quiescent period has no counterpart in the SMC cluster
  distribution (which has been steadily increasing with time).

\section{Summary and conclusions}

This paper is the first of a series devoted to quantitative
reconstruction of the SMC SFH from the deep HST/ACS observations
presented in \cite{sabbi09}. Here we have explored the directions SFH1
and SFH4, both located in the SMC's Bar, by comparing the
observational CMDs with a library of model CMDs incorporating
photometric uncertainties and incompleteness as estimated by
\cite{sabbi09}. To provide a robust characterization of the SFH, the
choice of the best model CMD was independently conducted with two
objective statistical methods, namely Cole's \citep{cole07} and
Bologna's \citep{cignoni10} procedures.

Our best simulated CMDs exhibit an overall good agreement with
observational CMDs. The star counts along the MS and the SGB
morphology are generally well reproduced, indicating that our
recovered SFHs and metallicity distributions are reasonable. However,
while SFH4 CMD is well fitted, there are some difficulties to
reproduce the exact morphology of SFH1's CMD, especially the upper-MS
spread and RC counts, which are underestimated and overestimated
respectively.  Concerning the resulting SFHs, a good consistency is
found between the two methods in both fields. The only significant
difference is the stronger rate suggested by Cole's SFH4 solution
between 1.5 and 3~Gyr ago.

  The combination of synthetic CMDs which most resembles the
  observations suggests the following picture. At early times, both
fields experienced a long quiescent phase characterized by low SFRs,
followed by a rapid SF increase around 5-6 Gyr ago. Since then, the
mode of star formation has been somewhat different in the two
fields. In SFH1 the star formation was gasping and reasonably high up
to today. In SFH4 it was smoother and slowly declining.

To account for these differences and similarities possible
explanations are:

\emph{Recent burstiness:} The different level of burstiness is not
surprising because these fields are separated by a distance (850~pc)
larger than most of the star forming complexes discovered in the SMC
(see \citealt{livanou2007}), thus allowing the recent activity to
fluctuate independently;

\emph{Recent systematic behavior:} The stellar density in the SFH4
region is lower than in SFH1. Hence the systematic decrease of SF
activity in SFH4 may be connected with a minor amount of fuel
available to support it up to today;


\emph{Early quiescence and prompt rise:} Are the quiescent period and
rapid SF increase 5-6 Gyr ago a global property of the Bar?  Recent
observations of RGB stars have revealed that older stellar components
of the SMC have a velocity dispersion of about 27.5 km s$^{-1}$
\citep{harris06}, high enough to distribute the stars over a large
distance (of the order of few kpc) from their birth places within few
Gyr. Hence, the low early activity and the prompt rise are not
peculiarities of our fields but global features of the SMC
Bar. Moreover, \cite{subra2012} find no evidence for a Bar in older
stars, which is consistent with our low early activity. Further
support is also provided by \cite{chiosi07}, who found similar star
formation trends for three other Bar fields located around the SMC
clusters K~29, NGC~290, and NGC~265. From the theoretical point of
view it is not clear what mechanism is responsible for the rapid rise
of stellar production. Was it externally triggered by the MW or the
LMC, or self-initiated? The striking similarity between the SMC and
LMC SFH is suggestive of the former. Pointing in this direction are
the recent calculations by \cite{diaz2011} who presented evidence that
around 5.5 Gyr ago the LMC and SMC were within 160 kpc of the MW and
200 kpc of each other, therefore arguing for an independent origin of
the Clouds. In this scenario the transition between quiescent and
active phases could be naturally explained in terms of growing rate of
mutual interactions started around 5 Gyr ago.

Finally, \cite{rafe2005} noted that the spatial
distribution of the younger and older clusters in the SMC is
statistically different, leading to the inference that a significant
accretion or merger event may have taken place around 3$-$5 Gyr ago.

The study reported in this paper was the first step in a wider
research activity aimed to characterize the SMC SFH through deep
HST/ACS observations. A forthcoming paper will be dedicated to the
analysis of the Wing and Halo fields. This will allow a comparative
analysis to look for global physical characteristics in the SMC star
formation process.

\acknowledgments MC and MT have been partially funded with contracts
COFIS ASI-INAF I/016/07/0 and ASI-INAF I/009/10/0. EKG acknowledges
partial funding from Sonderforschungsbereich "The Milky Way System"
(SFB 881) of the German Research Foundation (DFG), especially via
subproject A2. Partial support for JSG's analysis of data from HST
program GO-10396 was provided by NASA through a grant from the Space
Telescope Science Institute, which is operated by the Association of
Universities for Research in Astronomy, Inc., under NASA contract NAS
5-26555.





\clearpage

\end{document}